\documentclass[prd,nopacs,preprintnumbers,twocolumn,amsmath,nofootinbib,amssymb]{revtex4}
\usepackage{graphicx,color,dcolumn,booktabs,bm}
\usepackage{longtable,lscape}
\usepackage{txfonts}
\usepackage{overpic}
\usepackage{epstopdf}
\usepackage{indentfirst}
\usepackage{threeparttable}
\usepackage{feynmf}   
\usepackage{slashed}  
\usepackage{cases}
\usepackage{xcolor}
\usepackage{float}
\usepackage{makecell}
\usepackage{multirow}
\usepackage{array}
\usepackage{gensymb}
\usepackage{epsfig,dsfont,amssymb,amsmath,amsfonts,amsbsy,mathrsfs}

\graphicspath{{Figures/}} %

\usepackage{hyperref}
\hypersetup{colorlinks,citecolor=blue,anchorcolor=red,menucolor=red, linkcolor=red,filecolor=red,runcolor=red,urlcolor=blue,frenchlinks=red}

\makeatletter
\@addtoreset{equation}{section}
\makeatother

\allowdisplaybreaks

\newcolumntype{I}{!{\vrule width 0.9pt}}

\begin{document}

\title{The nonleptonic decays $\Xi_{cc}^{++}\to\Xi_{c}^{(\prime)+}\pi^{+}$ within the nonrelativistic quark model}
\author{Yu-Shuai Li$^{1}$}\email{liysh@pku.edu.cn}
\affiliation{$^1$ School of Physics and Center of High Energy Physics, Peking University, Beijing 100871, China}

\begin{abstract}

In this work, we study the nonleptonic decays $\Xi_{cc}^{++}\to\Xi_{c}^{(\prime)+}\pi^{+}$ with considering $\Xi_{c}-\Xi_{c}^{\prime}$ mixing. The relevant decay amplitudes are evaluated within the framework of nonrelativistic quark model, combining the baryon spatial wave functions adopted from solving the Schr\"{o}dinger equation with a nonrelativistic potential. With the mixing angle ranging $\theta\in(-18.2^{\circ},-14.3^{\circ})$, we successfully reproduce the measured ratio $R=\mathcal{B}[\Xi_{cc}^{++}\to\Xi_{c}^{\prime+}\pi^{+}]/\mathcal{B}[\Xi_{cc}^{++}\to\Xi_{c}^{+}\pi^{+}]$ reported by the LHCb Collaboration. Furthermore, we estimate the branching fractions as $\mathcal{B}[\Xi_{cc}^{++}\to\Xi_{c}^{+}\pi^{+}]=(3.2\sim4.3)\%$ and $\mathcal{B}[\Xi_{cc}^{++}\to\Xi_{c}^{\prime+}\pi^{+}]=(4.2\sim6.0)\%$, and the asymmetry parameters as $\alpha[\Xi_{cc}^{++}\to\Xi_{c}^{+}\pi^{+}]=(-0.80\sim-0.81)$ and $\alpha[\Xi_{cc}^{++}\to\Xi_{c}^{\prime+}\pi^{+}]=(-0.61\sim-0.62)$. The measurements of absolute branching fractions and asymmetry parameters by the ongoing LHCb and Belle II experiments will be helpful for further testing our numerical results and confirming the mixing angle.

\end{abstract}

\maketitle

\section{introduction}
\label{sec1}

\begin{figure*}[htbp]\centering
  \includegraphics[width=140mm]{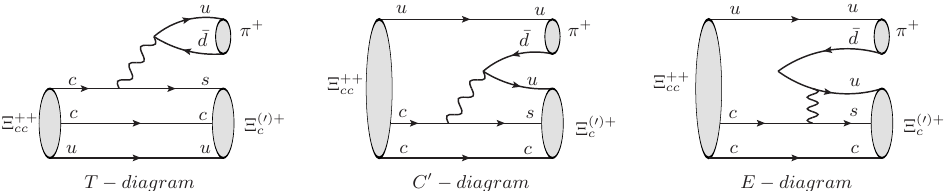}\\
  \caption{The topological diagrams for $\Xi_{cc}^{++}\to\Xi_{c}^{(\prime)+}\pi^{+}$ processes, in which the external $W$-emission diagram ($T$) is factorizable, while the inner $W$-emission diagram ($C^{\prime}$) and the $W$-exchange diagram ($E$) are nonfactorizable.}
  \label{fig:TopologicalDiagrams}
\end{figure*}

The study of charmed baryon nonleptonic decays provides valuable insights into weak decay mechanisms. At the tree level, the decay amplitudes are dominated by four topological diagrams, i.e., the external $W$-emission diagrams ($T$) and internal $W$-emission diagram ($C$), which are factorizable, and the inner $W$-emission diagram ($C^{\prime}$) and $W$-exchange diagram ($E$), which are nonfactorizable~\cite{Cheng:2021qpd}. Particularly in $\Xi_{cc}^{++}\to\Xi_{c}^{(\prime)+}\pi^{+}$ decays, there are factorizable $T$ diagram, and nonfactorizable $C^{\prime}$ and $E$ diagrams, as shown in Fig.~\ref{fig:TopologicalDiagrams}, contributing. However, due to the inherent nonperturbative nature of Quantum Chromodynamics (QCD), theoretical study remains challenging and incomplete.

As of now, the experimental measurement of absolute branching fractions of $\Xi_{cc}^{++}\to\Xi_{c}^{(\prime)+}\pi^{+}$ processes is absent. In 2022, the LHCb Collaboration reported the first observation of $\Xi_{cc}^{++}\to\Xi_{c}^{\prime+}\pi^{+}$ decay, and determined its branching fraction relating to $\Xi_{cc}^{++}\to\Xi_{c}^{+}\pi^{+}$ decay as~\cite{LHCb:2022rpd}
\begin{equation}
R=\frac{\mathcal{B}[\Xi_{cc}^{++}\to\Xi_{c}^{\prime+}\pi^{+}]}{\mathcal{B}[\Xi_{cc}^{++}\to\Xi_{c}^{+}\pi^{+}]}=1.41\pm0.17\pm0.10.
\end{equation}
When considering only factorizable amplitudes, theoretical studies including light-front quark model~\cite{Wang:2017mqp,Ke:2019lcf} and QCD sum rules~\cite{Shi:2019hbf} yielded $R<1$. A possible explanation of $R>1$ is to introduce the $\Xi_{c}-\Xi_{c}^{\prime}$ mixing effect~\cite{Ke:2022gxm}, i.e.,
\begin{equation}
\left(\begin{array}{c}
\Xi_{c}\\
\Xi_{c}^{\prime}
\end{array}\right)
=\left(\begin{array}{cc}
\cos{\theta} & \sin{\theta}\\
-\sin{\theta} & \cos{\theta}
\end{array}\right)
\left(\begin{array}{c}
\Xi_{c}^{\bar{3}}\\
\Xi_{c}^{6}
\end{array}\right)
\end{equation}
with $\Xi_{c}^{(\prime)}$ being the wave function of physical state, and $\Xi_{c}^{\bar{3}(6)}$ being that of antitriplet (sextet) state. As demonstrated in Ref.~\cite{Ke:2022gxm}, the ratio $R=0.56\pm0.18$ increases to $R=1.41\pm0.20$ when the mixing angle $\theta$ is either $(16.27\pm2.30)^{\circ}$ or $(85.54\pm2.30)^{\circ}$.

Apart from the mixing effect, the nonfactorizable contributions from $C^{\prime}$ and $E$ diagrams also need to be taken into account simultaneously, considering that the nonfactorizable amplitudes are comparable in magnitude to the factorizable amplitudes in charmed baryon weak decays~\cite{BESIII:2015bjk}. In theoretical study of $\Xi_{cc}$ weak decays, factorizable amplitudes can often be evaluated as the product of baryon transition matrix element and meson decay constant under the naive factorization assumption, while nonfactorizable amplitudes are more challenging to quantify. To address this challenge, several approaches have been developed, such as pole model~\cite{Sharma:2017txj,Cheng:2020wmk,Groote:2021pxt,Liu:2022igi,Zeng:2022egh,Liu:2023dvg}, light-cone sum rules (LCSR)~\cite{Shi:2022kfa}, and covariant confined quark model (CCQM)~\cite{Gutsche:2018msz,Ivanov:2020xmw}. In addition, methods not addressed in dynamics, including the irreducible SU(3) approach~\cite{Shi:2017dto,Wang:2017azm,Geng:2017mxn,Groote:2021pxt,Li:2021rfj,Wang:2022kwe} and topological diagram approach~\cite{Wang:2022wrb} based on SU(3)$_{\text{F}}$ symmetry in QCD, were also widely applied. Besides, the final state rescattering mechanism was also considered~\cite{Yu:2017zst,Li:2020qrh,Han:2021azw,Hu:2024uia}. In Refs.~\cite{Cheng:2020wmk,Liu:2022igi,Zeng:2022egh}, the authors employed pole model and soft-pion assumption to estimate nonfactorizable amplitudes, with nonparturbative parameters involved derived from bag model~\cite{Cheng:2020wmk,Liu:2022igi} and nonrelativistic quark model~\cite{Zeng:2022egh}, respectively. After considering the contributions from both mixing effects and nonfactorizable amplitudes, the measured ratio of branching fractions $R$ can be reproduced~\cite{Liu:2022igi,Zeng:2022egh}, but the asymmetry parameters show difference. The deviation is attributed to discrepancies between the baryon transition matrix elements predicted by different theoretical models. Further experimental determination of asymmetry parameters will be helpful for testing different works.

In the presented paper, we investigate the nonleptonic weak decays $\Xi_{cc}^{++}\to\Xi_{c}^{(\prime)+}\pi^{+}$ by utilizing a nonrelativistic quark model (NRQM) with considering $\Xi_{c}-\Xi_{c}^{\prime}$ mixing, in which particularly the nonfactorizable contributions from $W$-exchange diagrams are evaluated by pole model. We reiterate that in theoretical study of charmed baryon nonleptonic decays, the primary uncertainties stem from the baryon transition matrix element~\cite{Liu:2022igi,Li:2025alu}. Addressing this challenge, we propose using exact baryon wave functions, obtained by solving the Schr\"{o}dinger equation with a nonrelativistic potential, rather than relying on an oversimplified Gaussian-typed one. With the support from charmed baryon spectroscopy, our study minimizes the dependence on arbitrary wave functions for decay amplitudes, thereby reducing the corresponding uncertainties.

This paper is organized as follows: After the Introduction, we derive the weak decay amplitudes within the NRQM in Sec.~\ref{sec2}. And then, to obtain the involved baryon spatial wave functions, a nonrelativistic potential is introduced in Sec.~\ref{sec3}. In Sec.~\ref{sec4}, we present our numerical results of the decay amplitudes and future investigate the branching fraction and asymmetry parameter. Finally, this paper ends with a short summary in Sec.~\ref{sec5}.

\section{The decay amplitudes and physical observables}
\label{sec2}

\subsection{The decay amplitudes of $c\to su\bar{d}$ processes}
\label{sec2.1}

\begin{figure}[htbp]\centering
  \includegraphics[width=86mm]{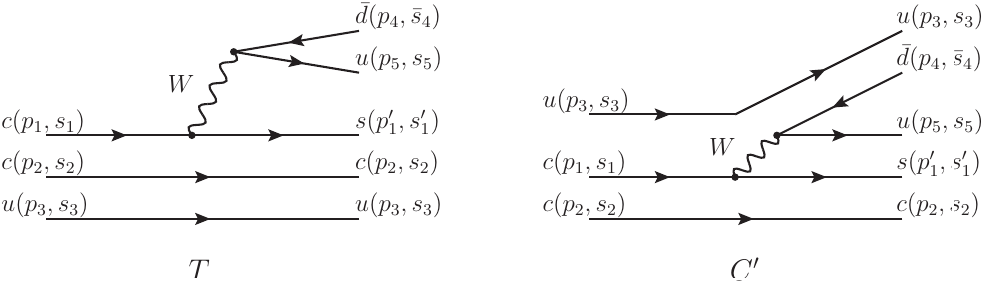}\\
  \caption{The kinematic conventions of external $W$-emission diagram ($T$) and the inner $W$-emission diagram ($C^{\prime}$) of $c\to su\bar{d}$ transition, where $p_{i}$ and $s_{i}$ are three-momentum and spin of $i$th quark, respectively.}
  \label{fig:fig13}
\end{figure}

According to the conventions in Fig.~\ref{fig:fig13}, the effective weak Hamiltonian of ${c}\to{su\bar{d}}$ transition at the tree level can be written as
\begin{equation}
\begin{split}
\mathcal{H}_{{c}\to{su\bar{d}}}=&\frac{G_{F}}{\sqrt{2}}V_{cs}V_{ud}\frac{\beta}{(2\pi)^{3}}\delta^{3}(\pmb{p}_{1}-\pmb{p}_{1}^{\prime}-\pmb{p}_{4}-\pmb{p}_{5})\\
&\times\bar{u}(\pmb{p}_{1}^{\prime},m_{1}^{\prime})\gamma_{\mu}(1-\gamma_{5})u(\pmb{p}_{1},m_{1})\\
&\times\bar{u}(\pmb{p}_{5},m_{5})\gamma^{\mu}(1-\gamma_{5})\nu(\pmb{p}_{4},m_{4})\hat{\alpha}_{1}^{(-)}\hat{I}_{\pi},
\end{split}
\end{equation}
where, $V_{cs}$ and $V_{ud}$ are the Cabibbo-Kobayashi-Maskawa (CKM) matrix elements; $\pmb{p}_{i}$ and $m_{i}$ denote the three-momentum and mass of $i$th quark, respectively; $\hat{\alpha}_{1}^{(-)}c=s$ represents the flavor-changing operator; and $\hat{I}_{\pi}=b_{d}^{\dagger}b_{u}$ represents the isospin operator for $\pi^{+}$ production, with $b_{d}^{\dagger}$ and $b_{u}$ denoting the creation and annihilation operators. Besides, the factor $\beta$ is assigned a value of 2 for the $T$ diagram, and $2/3$ for the $C^{\prime}$ diagram.

In the nonrelativistic limit, the effective Hamiltonian can be rewritten as~\cite{Niu:2020gjw,Niu:2021qcc,Niu:2025lgt}
\begin{widetext}
\begin{equation}
\begin{split}
\mathcal{H}_{{c}\to{su\bar{d}}}^{\text{PC}}=&\frac{G_{F}}{\sqrt{2}}V_{cs}V_{ud}\frac{\beta}{(2\pi)^{3}}\delta^{3}(\pmb{p}_{1}-\pmb{p}_{1}^{\prime}-\pmb{p}_{4}-\pmb{p}_{5})
\Bigg{\{}\langle{s_{z,1}^{\prime}}\vert{I}\vert{s_{z,1}}\rangle\langle{s_{z,5}\bar{s}_{z,4}}\vert{\pmb{\sigma}}\vert{0}\rangle\Bigg{(}\frac{\pmb{p}_{5}}{2m_{5}}+\frac{\pmb{p}_{4}}{2m_{4}}\Bigg{)}\\
&-\Bigg{[}\Bigg{(}\frac{\pmb{p}_{1}^{\prime}}{2m_{1}^{\prime}}+\frac{\pmb{p}_{1}}{2m_{1}}\Bigg{)}\langle{s_{z,1}^{\prime}}\vert{I}\vert{s_{z,1}}\rangle-i\langle{s_{z,1}^{\prime}}\vert{\pmb{\sigma}}\vert{s_{z,1}}\rangle\times\Bigg{(}\frac{\pmb{p}_{1}}{2m_{1}}-\frac{\pmb{p}_{1}^{\prime}}{2m_{1}^{\prime}}\Bigg{)}\Bigg{]}\langle{s_{z,5}\bar{s}_{z,4}}\vert{\pmb{\sigma}}\vert{0}\rangle\\
&-\langle{s_{z,1}^{\prime}}\vert{\pmb{\sigma}}\vert{s_{z,1}}\rangle\Bigg{[}\Bigg{(}\frac{\pmb{p}_{5}}{2m_{5}}+\frac{\pmb{p}_{4}}{2m_{4}}\Bigg{)}\langle{s_{z,5}\bar{s}_{z,4}}\vert{I}\vert{0}\rangle-i\langle{s_{z,5}\bar{s}_{z,4}}\vert{\pmb{\sigma}}\vert{0}\rangle\times\Bigg{(}\frac{\pmb{p}_{4}}{2m_{4}}-\frac{\pmb{p}_{5}}{2m_{5}}\Bigg{)}\Bigg{]}\\
&+\langle{s_{z,1}^{\prime}}\vert{\pmb{\sigma}}\vert{s_{z,1}}\rangle\Bigg{(}\frac{\pmb{p}_{1}^{\prime}}{2m_{1}^{\prime}}+\frac{\pmb{p}_{1}}{2m_{1}}\Bigg{)}\langle{s_{z,5}\bar{s}_{z,4}}\vert{I}\vert{0}\rangle\Bigg{\}}\hat{\alpha}_{1}^{(-)}\hat{I}_{\pi},\\
\mathcal{H}_{{c}\to{su\bar{d}}}^{\text{PV}}=&\frac{G_{F}}{\sqrt{2}}V_{cs}V_{ud}\frac{\beta}{(2\pi)^{3}}\delta^{3}(\pmb{p}_{1}-\pmb{p}_{1}^{\prime}-\pmb{p}_{4}-\pmb{p}_{5})
\Bigg{(}-\langle{s_{z,1}^{\prime}}\vert{I}\vert{s_{z,1}}\rangle\langle{s_{z,5}\bar{s}_{z,4}}\vert{I}\vert{0}\rangle+\langle{s_{z,1}^{\prime}}\vert{\pmb{\sigma}}\vert{s_{z,1}}\rangle\langle{s_{z,5}\bar{s}_{z,4}}\vert{\pmb{\sigma}}\vert{0}\rangle\Bigg{)}\hat{\alpha}_{1}^{(-)}\hat{I}_{\pi},
\label{eq:H13}
\end{split}
\end{equation}
\end{widetext}
in which, the parity-conserving (PC) Hamiltonian and parity-violating (PV) one are distinguished by odd or even number of $\gamma_{5}$, respectively. Besides, $I$ is the two-rank identity matrix, and $\pmb{\sigma}=(\sigma^{1},\sigma^{2},\sigma^{3})$ are the Pauli matrices. In addition, $\vert{s_{z,i}}\rangle$ is an abbreviation of $\vert{s_{i},s_{z,i}}\rangle$ and represent the spin of $i$th quark. {In charmed baryon weak decay, the relativistic effect may not be negligible. However, the precise calculation of relativistic effect is beyond the present work, we still work in the nonrelativistic framework.}

At the tree level, the amplitudes of $T$ and $C^{\prime}$ diagrams of $\mathcal{B}_{i}(\pmb{P}_{i})\to\mathcal{B}_{f}(\pmb{P}_{f})+P(\pmb{k})(P\equiv\text{pseudoscalar~meson})$ process can be evaluated by
\begin{equation}
\mathcal{M}_{T(C^{\prime}),\text{PC}(\text{PV})}^{J_{z}^{f},J_{z}^{i}}=\langle{\mathcal{B}_{f}(\pmb{P}_{f};J^{f},J_{z}^{f})P(\pmb{k})}\vert{\mathcal{H}_{{c}\to{su\bar{d}}}^{\text{PC}(\text{PV})}}\vert{\mathcal{B}_{i}(\pmb{P}_{i};J^{i},J_{z}^{i})}\rangle,
\end{equation}
where, $J^{i}=J^{f}=1/2$ are omitted. Besides, $\pmb{P}_{i}$ is the three-momentum of initial double charmed baryon, $\pmb{P}_{f}=\pmb{p}_{1}+\pmb{p}_{2}+\pmb{p}_{3}^{\prime}(\pmb{P}_{f}=\pmb{p}_{1}^{\prime}+\pmb{p}_{2}+\pmb{p}_{5})$ and $\pmb{k}=\pmb{p}_{4}+\pmb{p}_{5}~(\pmb{k}=\pmb{p}_{3}+\pmb{p}_{4})$ is the three-momenta of final charmed baryon and pion meson, respectively for $T$ $(C^{\prime})$ diagram. In the rest frame of initial state, we have $\pmb{P}_{i}=0$ and $\pmb{P}_{f}=-\pmb{k}$. Generally, the decay amplitude can be separated as
\begin{equation}
\begin{split}
\mathcal{M}_{T,C^{\prime}}^{J_{z}^{f},J_{z}^{i}}=&\langle{\mathcal{B}_{f}(\pmb{P}_{f};J^{f},J_{z}^{f})P(\pmb{k})}\vert{\mathcal{H}_{{c}\to{su\bar{d}}}}\vert{\mathcal{B}_{i}(\pmb{P}_{i};J^{i},J_{z}^{i})}\rangle\\
=&\sum_{S_{z}^{f},L_{z}^{f},S_{z}^{i},L_{z}^{i}}\langle{S^{f},S_{z}^{f};L^{f},L_{z}^{f}}\vert{J^{f},J_{z}^{f}}\rangle\langle{S^{i},S_{z}^{i};L^{i},L_{z}^{i}}\vert{J^{i},J_{z}^{i}}\rangle\\
&\times\langle\phi_{f}\chi_{f}^{S^{f},S_{z}^{f}};\phi_{P}\chi_{P}\vert\hat{\mathcal{O}}_{{c}\to{su\bar{d}}}^{\text{spin}}\hat{\mathcal{O}}_{{c}\to{su\bar{d}}}^{\text{flavor}}\vert\phi_{i}\chi_{i}^{S^{i},S_{z}^{i}}\rangle\\
&\times\langle\psi_{P}(\pmb{k})\psi_{L^{f},L_{z}^{f}}(\pmb{P}_{f})\vert{\hat{\mathcal{O}}_{{c}\to{su\bar{d}}}^{\text{spatial}}}\vert\psi_{L^{i},L_{z}^{i}}(\pmb{P}_{i})\rangle,
\label{eq:amplitudes1}
\end{split}
\end{equation}
where $\langle{S,S_{z};L,L_{z}}\vert{J,J_{z}}\rangle$ is the Clebsch-Gordan coefficient, while $\phi$, $\chi$ and $\psi$ are the flavor, spin and spatial wave functions, respectively.  The overlap of color wave functions is always 1, and is omitted here. In this work, the flavor and spin wave functions of $\pi^{+}$ meson are adopted as $\phi_{\pi}=u\bar{d}$ and $\chi_{\pi}=\frac{1}{\sqrt{2}}(\uparrow\downarrow-\downarrow\uparrow)$, respectively. For calculating the spin matrix element, the particle-hole conjugation $\langle{j,-m}\vert\to{(-1)^{j+m}}\vert{j,m}\rangle$ is useful. This relation makes the antiquark spin transform as $\langle\bar{\uparrow}\vert{\to}\vert{\downarrow}\rangle$ and $\langle\bar{\downarrow}\vert\to -\vert{\uparrow}\rangle$. Besides, the spin-flavor wave functions of concerned charmed baryons are adopted as~\cite{Perez-Marcial:1989sch}
\begin{equation}
\begin{split}
\vert\Xi_{cc}^{++}\rangle=&-\frac{1}{\sqrt{3}}\big{[}ccu\chi_{S}+(13)+(23)\big{]},\\
\vert\Xi_{c}^{\bar{3},+}\rangle=&\frac{1}{\sqrt{6}}\big{[}(usc-suc)\chi_{A}+(13)+(23)\big{]},\\
\vert\Xi_{c}^{6,+}\rangle=&\frac{1}{\sqrt{6}}\big{[}(usc+suc)\chi_{S}+(13)+(23)\big{]},
\end{split}
\end{equation}
where
\begin{equation}
\begin{split}
abc\chi_{S}^{\uparrow}=&\frac{1}{\sqrt{6}}(2a^{\uparrow}b^{\uparrow}c^{\downarrow}-a^{\uparrow}b^{\downarrow}c^{\uparrow}-a^{\downarrow}b^{\uparrow}c^{\uparrow}),\\
abc\chi_{A}^{\uparrow}=&\frac{1}{\sqrt{2}}(a^{\uparrow}b^{\downarrow}c^{\uparrow}-a^{\downarrow}b^{\uparrow}c^{\uparrow}),
\end{split}
\end{equation}
for the $\vert{S,S_{z}}\rangle=\vert{1/2,+1/2}\rangle$ case, while
\begin{equation}
\begin{split}
abc\chi_{S}^{\downarrow}=&\frac{1}{\sqrt{6}}(-2a^{\downarrow}b^{\downarrow}c^{\uparrow}+a^{\uparrow}b^{\downarrow}c^{\downarrow}+a^{\downarrow}b^{\uparrow}c^{\downarrow}),\\
abc\chi_{A}^{\downarrow}=&\frac{1}{\sqrt{2}}(a^{\uparrow}b^{\downarrow}c^{\downarrow}-a^{\downarrow}b^{\uparrow}c^{\downarrow}),
\end{split}
\end{equation}
for the $\vert{S,S_{z}}\rangle=\vert{1/2,-1/2}\rangle$ case.

In Tables~\ref{tab:Tfactor} and \ref{tab:Cprimfactor}, we present the spin-flavor matrix elements involved of PC and PV amplitudes for $T$ and $C^{\prime}$ diagrams, respectively. Especially for the $T$ diagram, we emphasize that the spin matrix elements associating with $\langle{s_{z,5}\bar{s}_{z,4}}\vert{\sigma_{x}}\vert{0}\rangle$, $\langle{s_{z,5}\bar{s}_{z,4}}\vert{\sigma_{y}}\vert{0}\rangle$ and $\langle{s_{z,5}\bar{s}_{z,4}}\vert{\sigma_{z}}\vert{0}\rangle$, are all vanishing.

\begin{table}[htbp]\centering
\caption{The spin-flavor matrix elements for PC (up panel) and PV (bottom panel) amplitudes of $T$ diagram of $\Xi_{cc}^{++}\to\Xi_{c}^{\bar{3},+}\pi^{+}$ ($\Xi_{cc}^{++}\to\Xi_{c}^{6,+}\pi^{+}$) process. For simplicity, the spin and flavor wave functions of $\pi^{+}$ meson is omitted here.}
\label{tab:Tfactor}
\renewcommand\arraystretch{1.05}
\begin{tabular*}{86mm}{c@{\extracolsep{\fill}}cc}
\toprule[1pt]
\toprule[0.5pt]
\specialrule{0em}{0.5pt}{0.5pt}
matrix elements &$\langle{\Xi_{c}^{\bar{3}(6),+}}{\downarrow}\vert\hat{\mathcal{O}}\vert{\Xi_{cc}^{++}}{\downarrow}\rangle$  &$\langle{\Xi_{c}^{\bar{3}(6),+}}{\uparrow}\vert\hat{\mathcal{O}}\vert{\Xi_{cc}^{++}}{\uparrow}\rangle$\\
\specialrule{0em}{0.5pt}{0.5pt}
\midrule[0.5pt]
$\langle{s_{z,1}^{\prime}}\vert{\sigma_{z}}\vert{s_{z,1}}\rangle\langle{s_{z,5}\bar{s}_{z,4}}\vert{I}\vert{0}\rangle$  &$\frac{1}{3\sqrt{3}}$ $(\frac{5}{9})$  &$-\frac{1}{3\sqrt{3}}$ $(-\frac{5}{9})$\\
\midrule[0.5pt]
$\langle{s_{z,1}^{\prime}}\vert{I}\vert{s_{z,1}}\rangle\langle{s_{z,5}\bar{s}_{z,4}}\vert{I}\vert{0}\rangle$  &$-\frac{1}{\sqrt{3}}$ $(-\frac{1}{3})$  &$-\frac{1}{\sqrt{3}}$ $(-\frac{1}{3})$\\
\bottomrule[0.5pt]
\bottomrule[1pt]
\end{tabular*}
\end{table}

\begin{table}[htbp]\centering
\caption{The spin-flavor matrix elements for PC (up panel) and PV (bottom panel) amplitudes of $C^{\prime}$ diagram of $\Xi_{cc}^{++}\to\Xi_{c}^{\bar{3},+}\pi^{+}$ ($\Xi_{cc}^{++}\to\Xi_{c}^{6,+}\pi^{+}$) process. For simplicity, the spin and flavor wave functions of $\pi^{+}$ meson is omitted here.}
\label{tab:Cprimfactor}
\renewcommand\arraystretch{1.05}
\begin{tabular*}{86mm}{c@{\extracolsep{\fill}}cc}
\toprule[1pt]
\toprule[0.5pt]
\specialrule{0em}{0.5pt}{0.5pt}
matrix elements &$\langle{\Xi_{c}^{\bar{3}(6),+}}{\downarrow}\vert\hat{\mathcal{O}}\vert{\Xi_{cc}^{++}}{\downarrow}\rangle$  &$\langle{\Xi_{c}^{\bar{3}(6),+}}{\uparrow}\vert\hat{\mathcal{O}}\vert{\Xi_{cc}^{++}}{\uparrow}\rangle$\\
\specialrule{0em}{0.5pt}{0.5pt}
\midrule[0.5pt]
$\langle{s_{z,1}^{\prime}}\vert{I}\vert{s_{z,1}}\rangle\langle{s_{z,5}\bar{s}_{z,4}}\vert{\sigma_{z}}\vert{0}\rangle$  &$\frac{1}{6\sqrt{3}}$ $(-\frac{1}{18})$  &$-\frac{1}{6\sqrt{3}}$ $(\frac{1}{18})$\\
$\langle{s_{z,1}^{\prime}}\vert{\sigma_{z}}\vert{s_{z,1}}\rangle\langle{s_{z,5}\bar{s}_{z,4}}\vert{I}\vert{0}\rangle$  &$\frac{1}{6\sqrt{3}}$ $(\frac{5}{18})$  &$-\frac{1}{6\sqrt{3}}$ $(-\frac{5}{18})$\\
$(\langle{s_{z,1}^{\prime}}\vert{\pmb{\sigma}}\vert{s_{z,1}}\rangle\times\langle{s_{z,5}\bar{s}_{z,4}}\vert{\pmb{\sigma}}\vert{0}\rangle)_{z}$  &$0$ $(0)$  &$0$ $(0)$\\
\midrule[0.5pt]
$\langle{s_{z,1}^{\prime}}\vert{I}\vert{s_{z,1}}\rangle\langle{s_{z,5}\bar{s}_{z,4}}\vert{I}\vert{0}\rangle$  &$-\frac{1}{2\sqrt{3}}$ $(-\frac{1}{6})$  &$-\frac{1}{2\sqrt{3}}$ $(-\frac{1}{6})$\\
$\langle{s_{z,1}^{\prime}}\vert{\sigma_{x}}\vert{s_{z,1}}\rangle\langle{s_{z,5}\bar{s}_{z,4}}\vert{\sigma_{x}}\vert{0}\rangle$  &$\frac{1}{6\sqrt{3}}$ $(-\frac{1}{18})$  &$\frac{1}{6\sqrt{3}}$ $(-\frac{1}{18})$\\
$\langle{s_{z,1}^{\prime}}\vert{\sigma_{y}}\vert{s_{z,1}}\rangle\langle{s_{z,5}\bar{s}_{z,4}}\vert{\sigma_{y}}\vert{0}\rangle$  &$\frac{1}{6\sqrt{3}}$ $(-\frac{1}{18})$  &$\frac{1}{6\sqrt{3}}$ $(-\frac{1}{18})$\\
$\langle{s_{z,1}^{\prime}}\vert{\sigma_{z}}\vert{s_{z,1}}\rangle\langle{s_{z,5}\bar{s}_{z,4}}\vert{\sigma_{z}}\vert{0}\rangle$  &$\frac{1}{6\sqrt{3}}$ $(-\frac{1}{18})$  &$\frac{1}{6\sqrt{3}}$ $(-\frac{1}{18})$\\
\bottomrule[0.5pt]
\bottomrule[1pt]
\end{tabular*}
\end{table}

After separating out the spin-flavor parts, the remaining spatial wave function convolutions can be expressed as
\begin{widetext}
\begin{equation}
\begin{split}
\mathcal{I}_{T}^{L^{f},L_{z}^{f};L^{i},L_{z}^{i}}=&\langle{\psi_{P}(\pmb{k})\psi_{L^{f},L_{z}^{f}}(\pmb{P}_{f})}\vert\hat{\mathcal{O}}_{{c}\to{su\bar{d}}}^{\text{spatial}}(\pmb{p}_{i})\vert{\psi_{L^{i},L_{z}^{i}}(\pmb{P}_{i})}\rangle\\
=&\int{d\pmb{p}_{1}d\pmb{p}_{2}d\pmb{p}_{3}d\pmb{p}_{1}^{\prime}d\pmb{p}_{4}d\pmb{p}_{5}}\psi^{*}_{P}(\pmb{p}_{5},\pmb{p}_{4})\psi_{L^{f},L_{z}^{f}}^{*}(\pmb{p}_{1}^{\prime},\pmb{p}_{3},\pmb{p}_{2})\psi_{L^{i},L_{z}^{i}}(\pmb{p}_{1},\pmb{p}_{2},\pmb{p}_{3})\mathcal{O}_{{c}\to{su\bar{d}}}^{\text{spatial}}(\pmb{p}_{i})\\
&\times\delta^{3}(\pmb{k}-\pmb{p}_{5}-\pmb{p}_{4})\delta^{3}(\pmb{P}_{f}-\pmb{p}_{1}^{\prime}-\pmb{p}_{2}-\pmb{p}_{3})\delta^{3}(\pmb{P}_{i}-\pmb{p}_{1}-\pmb{p}_{2}-\pmb{p}_{3})\delta^{3}(\pmb{p}_{1}-\pmb{p}_{4}-\pmb{p}_{5}-\pmb{p}_{1}^{\prime}),
\end{split}
\end{equation}
\begin{equation}
\begin{split}
\mathcal{I}_{C^{\prime}}^{L^{f},L_{z}^{f};L^{i},L_{z}^{i}}=&\langle{\psi_{P}(\pmb{k})\psi_{L^{f},L_{z}^{f}}(\pmb{P}_{f})}\vert\hat{\mathcal{O}}_{{c}\to{su\bar{d}}}^{\text{spatial}}(\pmb{p}_{i})\vert{\psi_{L^{i},L_{z}^{i}}(\pmb{P}_{i})}\rangle\\
=&\int{d\pmb{p}_{1}d\pmb{p}_{2}d\pmb{p}_{3}d\pmb{p}_{1}^{\prime}d\pmb{p}_{4}d\pmb{p}_{5}}\psi^{*}_{P}(\pmb{p}_{3},\pmb{p}_{4})\psi_{L^{f},L_{z}^{f}}^{*}(\pmb{p}_{1}^{\prime},\pmb{p}_{5},\pmb{p}_{2})\psi_{L^{i},L_{z}^{i}}(\pmb{p}_{1},\pmb{p}_{2},\pmb{p}_{3})\mathcal{O}_{{c}\to{su\bar{d}}}^{\text{spatial}}(\pmb{p}_{i})\\
&\times\delta^{3}(\pmb{k}-\pmb{p}_{3}-\pmb{p}_{4})\delta^{3}(\pmb{P}_{f}-\pmb{p}_{1}^{\prime}-\pmb{p}_{5}-\pmb{p}_{2})\delta^{3}(\pmb{P}_{i}-\pmb{p}_{1}-\pmb{p}_{2}-\pmb{p}_{3})\delta^{3}(\pmb{p}_{1}-\pmb{p}_{4}-\pmb{p}_{5}-\pmb{p}_{1}^{\prime}),
\end{split}
\end{equation}
\end{widetext}
where, $\mathcal{O}_{{c}\to{su\bar{d}}}^{\text{spatial}}(\pmb{p}_{i})$ is the function of $\pmb{p}_{i}/(2m_{i})$ for $\text{PC}$ amplitudes, and $\mathcal{O}_{{c}\to{su\bar{d}}}^{\text{spatial}}(\pmb{p}_{i})=1$ for $\text{PV}$ amplitudes.

After synthesizing the above discussion, obviously we have the following relations:
\begin{equation}
\begin{split}
\mathcal{M}_{\text{PC}}^{-1/2,-1/2}=&-\mathcal{M}_{\text{PC}}^{1/2,1/2},~\mathcal{M}_{\text{PV}}^{-1/2,-1/2}=\mathcal{M}_{\text{PV}}^{1/2,1/2},\\
\mathcal{M}_{\text{PV},\text{PC}}^{-1/2,1/2}=&\mathcal{M}_{\text{PV},\text{PC}}^{1/2,-1/2}=0,
\label{eq:relation1}
\end{split}
\end{equation}
for $T$ and $C^{\prime}$ diagrams.

To describe the spatial wave function of pion meson, the simple harmonic oscillator wave function~\cite{Niu:2020gjw,Niu:2021qcc}
\begin{equation}
\psi_{\pi}(\pmb{k}_{1},\pmb{k}_{2})=\frac{1}{\pi^{3/4}R^{3/2}}\exp\Big{[}-\frac{(\pmb{k}_{1}-\pmb{k}_{2})^{2}}{8R^{2}}\Big{]},
\end{equation}
is employed in our calculation, with the phenomenal parameter $R=0.28~\text{GeV}$~\cite{Niu:2020gjw,Niu:2021qcc}, while for that of charmed baryons, we adopt the expressions as
\begin{equation}
\begin{split}
\psi_{L,L_{z}}(\pmb{k}_{1},\pmb{k}_{2},\pmb{k}_{3})=&\sum_{n_{\rho},n_{\lambda}}\sum_{m_{\rho},m_{\lambda}}\langle{l_{\rho},m_{\rho};l_{\lambda},m_{\lambda}}\vert{L,L_{z}}\rangle\\
&\times{C^{(n_{\rho},n_{\lambda})}}\psi_{n_{\rho},l_{\rho},m_{\rho}}(\pmb{\rho})\psi_{n_{\lambda},l_{\lambda},m_{\lambda}}(\pmb{\lambda}),
\end{split}
\end{equation}
where the momentums of $\rho$-mode and $\lambda$-mode are defined as
\begin{equation}
\begin{split}
\pmb{\rho}=&\frac{m_{1}\pmb{k}_{2}-m_{2}\pmb{k}_{1}}{m_{1}+m_{2}},\\
\pmb{\lambda}=&\frac{(m_{1}+m_{2})\pmb{k}_{3}-m_{3}(\pmb{k}_{1}+\pmb{k}_{2})}{m_{1}+m_{2}+m_{3}},
\end{split}
\end{equation}
respectively. This treatment assumes that the single heavy baryon is treated as a bound state of a light quark cluster and a heavy quark, while the double heavy baryon is treated as a bound state of a heavy quark cluster and a light quark. Besides, $C^{(n_{\rho},n_{\lambda})}$ is a set of coefficients, and $\psi_{n,l,m}(\pmb{p})$ is the Gaussian basis function~\cite{Hiyama:2003cu}
\begin{equation}
\begin{split}
\psi_{n,l,m}(\pmb{p})=&N_{nl}~k^{l}~e^{-\nu_{n}p^{2}}Y_{lm}(\hat{\pmb{p}}),\\
N_{nl}=&\Bigg{(}\frac{2^{l+2}(2\nu_{n})^{l+3/2}}{\sqrt{\pi}(2l+1)!!}\Bigg{)}^{1/2},
\end{split}
\end{equation}
where $p\equiv\vert\pmb{p}\vert$, and $\hat{\pmb{p}}=\pmb{p}/p$ is a unit vector. $\nu_{n}$ is the Gaussian size parameter. The detailed discussions of spatial wave functions of baryons will be presented in Sec.~\ref{sec3}.

\subsection{The decay amplitudes of $cd\to su$ processes}
\label{sec2.2}

In this work, we adopt the pole model to estimate the amplitude of $W$-exchange diagram. The pole model has been successful applied to evaluate the nonfactorizable amplitudes of charmed baryons nonleptonic decays~\cite{Cheng:1991sn,Cheng:1992ff,Cheng:1993gf,Uppal:1994pt,Dhir:2018twm,Cheng:2018hwl,Zou:2019kzq,Hu:2020nkg,Cheng:2020wmk,Niu:2020gjw,Meng:2020euv,Niu:2021qcc,Liu:2022igi,Cheng:2022kea,Zeng:2022egh,Cheng:2022jbr,Ivanov:2023wir,Niu:2025lgt}. Within the framework of pole model, the general expression of decay amplitude of $\mathcal{B}_{i}(\pmb{P}_{i})\to\mathcal{B}_{f}(\pmb{P}_{f})+P(\pmb{k})$ process can be written as~\cite{Niu:2020gjw,Niu:2021qcc,Niu:2025lgt}
\begin{equation}
\begin{split}
i\mathcal{M}_{\mathcal{B}_{m}-pole}^{J_{z}^{f},J_{z}^{i}}=&\langle{\mathcal{B}_{f}(\pmb{P}_{f};J^{f},J_{z}^{f})}\vert{\mathcal{H}_{{cd}\to{su}}}\vert{\mathcal{B}_{m}(\pmb{P}_{f};J^{f},J_{z}^{f})}\rangle\\
&\times\frac{i}{\slashed{p}_{\mathcal{B}_{m}}-m_{\mathcal{B}_{m}}+i\Gamma_{\mathcal{B}_{m}}/2}\\
&\times\langle{\mathcal{B}_{m}(\pmb{P}_{f};J^{f},J_{z}^{f})}\vert{\mathcal{H}_{P}}\vert{\mathcal{B}_{i}(\pmb{P}_{i};J^{i},J_{z}^{i})}\rangle,
\label{eq:amplitudes2}
\end{split}
\end{equation}
where $\pmb{P}_{i}=0$, and $\pmb{P}_{f}=-\pmb{k}$ with $\pmb{k}=(0,0,k)$, and $\mathcal{B}_{m}$ represent the intermediate baryon states with quantum numbers of $J^{P}=1/2^{+}$ and $1/2^{-}$. In our calculation, the approximation of propagator
\begin{equation}
\frac{1}{\slashed{p}-m+i\Gamma/2}\approx\frac{2m}{p^{2}-m^{2}+i\Gamma m},
\end{equation}
is applied. Moreover, considering that the intermediate double charmed baryons involved are narrow~\cite{Xiao:2017dly,Xiao:2017udy}, and $m_{\mathcal{B}_{m}}^{2}-m_{\Xi_{c}^{(\prime)}}^{2}\gg\Gamma_{\mathcal{B}_{m}}m_{\mathcal{B}_{m}}$, we neglect the contribution from $i{\Gamma}{m}$ term in numerical calculation. In principle, all conceivable intermediate baryon states must be accounted for here. However as discussed in Refs.~\cite{Cheng:1991sn,Niu:2020gjw,Niu:2021qcc,Niu:2025lgt}, under the pole approximation, one usually concentrates on the most important low-lying $J^{P}=1/2^{+}$ and $1/2^{-}$ pole states. The nonfactorizable PV amplitudes primarily arise from low-lying $J^{P}=1/2^{-}$ poles, while the nonfactorizable PC amplitudes are determined by $J^{P}=1/2^{+}$ poles. Particularly for the concerned two-body nonleptonic weak decays $\Xi_{cc}^{++}\to\Xi_{c}^{(\prime),+}\pi^{+}$, we consider the intermediate states $\Xi_{cc}^{+}(1S)$ and $\Xi_{cc}^{+}(2S)$, who have the spin-parity quantum numbers $J^{P}=1/2^{+}$, and $\Xi_{cc}^{+}(1P,\rho)$, $\Xi_{cc}^{+}(1P,\lambda)$, $\Xi_{cc}^{+}(2P,\rho)$ and $\Xi_{cc}^{+}(2P,\lambda)$, who have the $J^{P}=1/2^{-}$. Here, $\Xi_{cc}^{+}(1S)$ and $\Xi_{cc}^{+}(2S)$ represent the ground state and the radial excited state, respectively; $\Xi_{cc}^{+}(1P,\rho)$ and $\Xi_{cc}^{+}(2P,\rho)$, and $\Xi_{cc}^{+}(1P,\lambda)$ and $\Xi_{cc}^{+}(2P,\lambda)$ represent the $\rho$-mode and $\lambda$-mode excited $P$-wave states, respectively.

\begin{figure}[htbp]\centering
  \includegraphics[width=50mm]{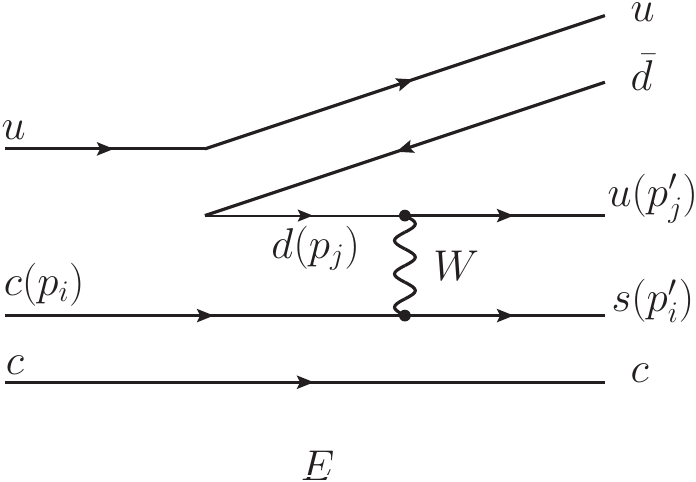}\\
  \caption{The kinematic conventions of the $W$-exchange diagram ($E$) of $cd\to su$ transition.}
  \label{fig:fig22}
\end{figure}

For the $cd\to su$ transition at the tree level, according to the conventions in Fig.~\ref{fig:fig22}, the effective weak Hamiltonian can be written as
\begin{equation}
\begin{split}
\mathcal{H}_{{cd}\to{su}}=&\frac{G_{F}}{\sqrt{2}}V_{cs}V_{ud}\frac{1}{(2\pi)^{3}}\delta^{3}(\pmb{p}_{i}^{\prime}+\pmb{p}_{j}^{\prime}-\pmb{p}_{i}-\pmb{p}_{j})\\
&\times\bar{u}(\pmb{p}_{i}^{\prime},m_{i}^{\prime})\gamma_{\mu}(1-\gamma_{5})u(\pmb{p}_{i},m_{i})\\
&\times\bar{u}(\pmb{p}_{j}^{\prime},m_{j}^{\prime})\gamma^{\mu}(1-\gamma_{5})u(\pmb{p}_{j},m_{j}).
\end{split}
\end{equation}
In the nonrelativistic limit, it can be rewritten as~\cite{Niu:2020gjw,Niu:2021qcc,Niu:2025lgt}
\begin{widetext}
\begin{equation}
\begin{split}
\mathcal{H}_{{cd}\to{su}}^{\text{PC}}=&\frac{G_{F}}{\sqrt{2}}V_{cs}V_{ud}\frac{1}{(2\pi)^{3}}\sum_{i{\neq}j}\hat{\alpha}_{i}^{(-)}\hat{\beta}_{j}^{(+)}\delta^{3}(\pmb{p}_{i}^{\prime}+\pmb{p}_{j}^{\prime}-\pmb{p}_{i}-\pmb{p}_{j})(1-\langle{s_{z,i}^{\prime}}\vert{\pmb{\sigma}_{i}}\vert{s_{z,i}}\rangle\langle{s_{z,j}^{\prime}}\vert{\pmb{\sigma}_{j}}\vert{s_{z,j}}\rangle),\\
\mathcal{H}_{{cd}\to{su}}^{\text{PV}}=&\frac{G_{F}}{\sqrt{2}}V_{cs}V_{ud}\frac{1}{(2\pi)^{3}}\sum_{i{\neq}j}\hat{\alpha}_{i}^{(-)}\hat{\beta}_{j}^{(+)}\delta^{3}(\pmb{p}_{i}^{\prime}+\pmb{p}_{j}^{\prime}-\pmb{p}_{i}-\pmb{p}_{j})\\
&\times\Bigg{\{}-\big{(}\langle{s_{z,i}^{\prime}}\vert{\pmb{\sigma}_{i}}\vert{s_{z,i}}\rangle-\langle{s_{z,j}^{\prime}}\vert{\pmb{\sigma}_{j}}\vert{s_{z,j}}\rangle\big{)}\Bigg{[}\Bigg{(}\frac{\pmb{p}_{i}}{2m_{i}}-\frac{\pmb{p}_{j}}{2m_{j}}\Bigg{)}+\Bigg{(}\frac{\pmb{p}_{i}^{\prime}}{2m_{i}^{\prime}}-\frac{\pmb{p}_{j}^{\prime}}{2m_{j}^{\prime}}\Bigg{)}\Bigg{]}\\
&+i\big{(}\langle{s_{z,i}^{\prime}}\vert{\pmb{\sigma}_{i}}\vert{s_{z,i}}\rangle\times\langle{s_{z,j}^{\prime}}\vert{\pmb{\sigma}_{j}}\vert{s_{z,j}}\rangle\big{)}\Bigg{[}\Bigg{(}\frac{\pmb{p}_{i}}{2m_{i}}-\frac{\pmb{p}_{j}}{2m_{j}}\Bigg{)}-\Bigg{(}\frac{\pmb{p}_{i}^{\prime}}{2m_{i}^{\prime}}-\frac{\pmb{p}_{j}^{\prime}}{2m_{j}^{\prime}}\Bigg{)}\Bigg{]}
\Bigg{\}},
\label{eq:Hamilton22}
\end{split}
\end{equation}
\end{widetext}
where, $\hat{\alpha}_{i}^{(-)}c=s$ and $\hat{\beta}_{j}^{(+)}d=u$ are the flavor-changing operators, which act on $i$th and $j$th quarks respectively.

Besides, the Hamiltonian $\mathcal{H}_{P}$ is adopted as quark-pseudoscalar meson coupling as
\begin{equation}
\mathcal{H}_{P}=\sum_{j}{\int}d{\pmb{x}}\frac{1}{f_{P}}\bar{q}_{j}(\pmb{x})\gamma_{\mu}\gamma_{5}q_{j}(\pmb{x})\partial^{\mu}\phi_{P}(\pmb{x}),
\end{equation}
where, $f_{P}$ is the pseudoscalar meson decay constant, and $q(\pmb{x})$ and $\phi_{P}$ denote the quark field and meson field respectively. In the nonrelativistic limit, specially for $\pi^{+}$ emitted process, the above effective Hamilton can be expanded as~\cite{Niu:2020gjw,Niu:2021qcc,Niu:2025lgt}
\begin{equation}
\begin{split}
\mathcal{H}_{\pi}=&\frac{1}{\sqrt{(2\pi)^{3}2\omega}}\sum_{j}\frac{1}{f_{\pi}}\Bigg{[}\omega\bigg{(}\frac{\pmb{\sigma}\cdot\pmb{p}_{f}^{j}}{2m_{f}}+\frac{\pmb{\sigma}\cdot\pmb{p}_{i}^{j}}{2m_{i}}\bigg{)}-\pmb{\sigma}\cdot\pmb{k}\Bigg{]}\\
&\times \hat{I}_{\pi}\delta^{3}(\pmb{p}_{f}^{j}+\pmb{k}-\pmb{p}_{i}^{j}),
\end{split}
\end{equation}
where, $\omega$ and $\pmb{k}$ are the energy and three-momentum of the pion meson in the rest frame of the initial state, respectively. $f_{\pi}=130~\text{MeV}$ is the decay constant of pion. Besides, $\pmb{p}_{i}^{j}$ and $\pmb{p}_{f}^{j}$ are the initial and final momentum of the $j$th quark, respectively. $\hat{I}_{\pi}=b_{d}^{\dagger}b_{u}$ is the isospin operator for the $\pi^{+}$ production with $b_{d}^{\dagger}$ and $b_{u}$ representing the creation and annihilation operators for quarks.

With the Hamiltonian in Eq.~\eqref{eq:Hamilton22}, the baryon weak transition matrix element of PC amplitude can be obtained by producing the integration as
\begin{widetext}
\begin{equation}
\begin{split}
\langle{\mathcal{B}_{f}(\pmb{P}_{f};J^{f},J_{z}^{f})}\vert{\mathcal{H}_{cd{\to}su}^{\text{PC}}}\vert{\mathcal{B}_{m}(\pmb{P}_{f};J^{f},J_{z}^{f})}\rangle=&
\frac{G_{F}}{\sqrt{2}}V_{cs}V_{ud}\frac{6}{(2\pi)^{3}}\int{d\pmb{p}_{1}}{d\pmb{p}_{2}}{d\pmb{p}_{3}}{d\pmb{p}_{1}^{\prime}}{d\pmb{p}_{2}^{\prime}}{d\pmb{p}_{3}^{\prime}}\delta^{3}(\pmb{p}_{1}^{\prime}+\pmb{p}_{2}^{\prime}-\pmb{p}_{1}-\pmb{p}_{2})\delta^{3}(\pmb{p}_{3}^{\prime}-\pmb{p}_{3})\\
&\times\sum_{S_{z}^{f},S_{z}^{m}}\sum_{L_{z}^{f},L_{z}^{m}}\langle{S^{f},S_{z}^{f};L^{f},L_{z}^{f}}\vert{J^{f},J_{z}^{f}}\rangle\langle{S^{m},S_{z}^{m};L^{m},L_{z}^{m}}\vert{J^{f},J_{z}^{f}}\rangle\\
&\times\langle{\mathcal{B}_{f}(S^{f},S_{z}^{f})}\vert\hat{\alpha}_{1}^{(-)}\hat{\beta}_{2}^{(+)}(1-\pmb{\sigma}_{1}\cdot\pmb{\sigma}_{2})\vert{\mathcal{B}_{m}(S^{m},S_{z}^{m})}\rangle\\
&\times\psi_{L^{f},L_{z}^{f}}^{*}(\pmb{p}_{1}^{\prime},\pmb{p}_{2}^{\prime},\pmb{p}_{3}^{\prime})\psi_{L^{m},L_{z}^{m}}(\pmb{p}_{1},\pmb{p}_{3},\pmb{p}_{2})
\delta^{3}(\pmb{P}_{f}-\pmb{p}_{1}-\pmb{p}_{2}-\pmb{p}_{3})\delta^{3}(\pmb{P}_{f}-\pmb{p}_{1}^{\prime}-\pmb{p}_{2}^{\prime}-\pmb{p}_{3}^{\prime}),
\end{split}
\end{equation}
\end{widetext}
where $6$ is a symmetry factor. It should be mention that the spin operator $(1-\pmb{\sigma}_{1}\cdot\pmb{\sigma}_{2})$ requires $S_{z}^{f}=S_{z}^{m}$. To deal with the spin-flavor matrix element, one can expand $\pmb{\sigma}_{1}\cdot\pmb{\sigma}_{2}$ as
\begin{equation}
\pmb{\sigma}_{1}\cdot\pmb{\sigma}_{2}=\frac{1}{2}\Big{(}\sigma_{1+}\sigma_{2-}+\sigma_{1-}\sigma_{2+}\Big{)}+\sigma_{1z}\sigma_{2z},
\end{equation}
where $\sigma_{\pm}=\sigma_{x}{\pm}{i\sigma_{y}}$, and the subscripts 1 and 2 denote that the spin operators act on the first and second quarks, respectively. Particularly, the spin-flavor matrix elements of PC amplitudes are presented in Table~\ref{tab:polefactor}.

\begin{table}[htbp]\centering
\caption{The spin-flavor matrix elements for PC (up panel) and PV (bottom panel) amplitudes of $W$-exchange diagrams of $\Xi_{cc}^{++}\to\Xi_{c}^{\bar{3},+}\pi^{+}$ ($\Xi_{cc}^{++}\to\Xi_{c}^{6,+}\pi^{+}$) process.}
\label{tab:polefactor}
\renewcommand\arraystretch{1.05}
\begin{tabular*}{86mm}{c@{\extracolsep{\fill}}cc}
\toprule[1pt]
\toprule[0.5pt]
\specialrule{0em}{0.5pt}{0.5pt}
matrix elements &$\langle{\Xi_{c}^{\bar{3}(6),+}}{\downarrow}\vert\hat{\mathcal{O}}\vert{\Xi_{cc}^{+}}{\downarrow}\rangle$  &$\langle{\Xi_{c}^{\bar{3}(6),+}}{\uparrow}\vert\hat{\mathcal{O}}\vert{\Xi_{cc}^{+}}{\uparrow}\rangle$\\
\specialrule{0em}{0.5pt}{0.5pt}
\midrule[0.5pt]
$\hat{\alpha}_{1}^{(-)}\hat{\beta}_{2}^{(+)}(1-\pmb{\sigma}_{1}\cdot\pmb{\sigma}_{2})$  &$\sqrt{\frac{2}{3}}$ $(0)$  &$\sqrt{\frac{2}{3}}$ $(0)$\\
\midrule[0.5pt]
$-\hat{\alpha}_{1}^{(-)}\hat{\beta}_{2}^{(+)}(\sigma_{1z}-\sigma_{2z})$  &$\frac{1}{3\sqrt{6}}$ $(\frac{1}{3\sqrt{2}})$  &$-\frac{1}{3\sqrt{6}}$ $(-\frac{1}{3\sqrt{2}})$\\
$\hat{\alpha}_{1}^{(-)}\hat{\beta}_{2}^{(+)}(\pmb{\sigma}_{1}\times\pmb{\sigma}_{2})_{z}$  &$-\frac{i}{3\sqrt{6}}$ $(\frac{i}{3\sqrt{2}})$  &$\frac{i}{3\sqrt{6}}$ $(-\frac{i}{3\sqrt{2}})$\\
\bottomrule[0.5pt]
\bottomrule[1pt]
\end{tabular*}
\end{table}

Analogously, the baryon weak transition matrix element of PV amplitude can be obtained by producing the integration as
\begin{widetext}
\begin{equation}
\begin{split}
\langle{\mathcal{B}_{f}(\pmb{P}_{f};J^{f},J_{z}^{f})}\vert{\mathcal{H}_{cd{\to}su}^{\text{PV}}}\vert{\mathcal{B}_{m}(\pmb{P}_{f};J^{f},J_{z}^{f})}\rangle=&
\frac{G_{F}}{\sqrt{2}}V_{cs}V_{ud}\frac{6}{(2\pi)^{3}}\int{d\pmb{p}_{1}}{d\pmb{p}_{2}}{d\pmb{p}_{3}}{d\pmb{p}_{1}^{\prime}}{d\pmb{p}_{2}^{\prime}}{d\pmb{p}_{3}^{\prime}}\delta^{3}(\pmb{p}_{1}^{\prime}+\pmb{p}_{2}^{\prime}-\pmb{p}_{1}-\pmb{p}_{2})\delta^{3}(\pmb{p}_{3}^{\prime}-\pmb{p}_{3})\\
&\times\sum_{S_{z}^{f},S_{z}^{m}}\sum_{L_{z}^{f},L_{z}^{m}}\langle{S^{f},S_{z}^{f};L^{f},L_{z}^{f}}\vert{J^{f},J_{z}^{f}}\rangle\langle{S^{m},S_{z}^{m};L^{m},L_{z}^{m}}\vert{J^{f},J_{z}^{f}}\rangle\\
&\times\langle{\mathcal{B}_{f}(S^{f},S_{z}^{f})}\vert\hat{\alpha}_{1}^{(-)}\hat{\beta}_{2}^{(+)}\hat{\mathcal{O}}_{cd{\to}su}^{\text{spin}}\vert{\mathcal{B}_{m}(S^{m},S_{z}^{m})}\rangle\\
&\times\psi_{L^{f},L_{z}^{f}}^{*}(\pmb{p}_{1}^{\prime},\pmb{p}_{2}^{\prime},\pmb{p}_{3}^{\prime})\mathcal{O}_{cd{\to}su}^{\text{spatial}}(\pmb{p}_{i})\psi_{L^{m},L_{z}^{m}}(\pmb{p}_{1},\pmb{p}_{2},\pmb{p}_{3}),
\end{split}
\end{equation}
where $\mathcal{O}_{cd{\to}su}^{\text{spatial}}(\pmb{p}_{i})$ is the function of $\pmb{p}_{i}/(2m_{i})$. Moreover, the concerned spin-flavor matrix elements of PV amplitudes are also presented in Table~\ref{tab:polefactor}.

As for the transition matrix element $\langle{\mathcal{B}_{m}(\pmb{P}_{f})}\vert{\mathcal{H}_{\pi}}\vert{\mathcal{B}_{i}(\pmb{P}_{i})}\rangle$, we have
\begin{equation}
\begin{split}
\langle{\mathcal{B}_{m}(\pmb{P}_{f};J^{f},J_{z}^{f})}\vert{\mathcal{H}_{\pi}}\vert{\mathcal{B}_{i}(\pmb{P}_{i};J^{i},J_{z}^{i})}\rangle=&\frac{1}{\sqrt{(2\pi)^{3}2\omega}}\frac{1}{f_{\pi}}
\int{d\pmb{p}_{1}}{d\pmb{p}_{2}}{d\pmb{p}_{3}}{d\pmb{p}_{1}^{\prime}}{d\pmb{p}_{2}^{\prime}}{d\pmb{p}_{3}^{\prime}}\delta^{3}(\pmb{p}_{1}^{\prime}-\pmb{p}_{1})\delta^{3}(\pmb{p}_{2}^{\prime}-\pmb{p}_{2})\delta^{3}(\pmb{p}_{3}^{\prime}-\pmb{p}_{3}+\pmb{k})\\
&\times\sum_{S_{z}^{i},S_{z}^{m}}\sum_{L_{z}^{i},L_{z}^{m}}\langle{S^{i},S_{z}^{i};L^{i},L_{z}^{i}}\vert{J^{i},J_{z}^{i}}\rangle\langle{S^{m},S_{z}^{m};L^{m},L_{z}^{m}}\vert{J^{f},J_{z}^{f}}\rangle\\
&\times\langle{\mathcal{B}_{m}(S^{m},S_{z}^{m})}\vert\hat{I}_{\pi}(-\pmb{\sigma}\cdot\pmb{k})\vert{\mathcal{B}_{i}(S^{i},S_{z}^{i})}\rangle{\frac{\omega+4m_{c}+2m_{n}}{2(2m_{c}+m_{n})}}\\
&\times\psi_{L^{m},L_{z}^{m}}^{*}(\pmb{p}_{1}^{\prime},\pmb{p}_{2}^{\prime},\pmb{p}_{3}^{\prime})\psi_{L^{i},L_{z}^{i}}(\pmb{p}_{1},\pmb{p}_{2},\pmb{p}_{3})
\delta^{3}(\pmb{P}_{i}-\pmb{p}_{1}-\pmb{p}_{2}-\pmb{p}_{3})\delta^{3}(\pmb{P}_{f}-\pmb{p}_{1}^{\prime}-\pmb{p}_{2}^{\prime}-\pmb{p}_{3}^{\prime}),
\end{split}
\end{equation}
\end{widetext}
in which, the non-vanishing spin-flavor matrix elements show
\begin{equation}
\begin{split}
\langle{\Xi_{cc}^{+}{\downarrow}}\vert{\hat{I}_{\pi}}(-\pmb{\sigma}\cdot\pmb{k})\vert{\Xi_{cc}^{++}{\downarrow}}\rangle=&k,\\
\langle{\Xi_{cc}^{+}{\uparrow}}\vert{\hat{I}_{\pi}}(-\pmb{\sigma}\cdot\pmb{k})\vert{\Xi_{cc}^{++}{\uparrow}}\rangle=-k&.
\end{split}
\end{equation}

After synthesizing the above discussion, obviously we have the relations as
\begin{equation}
\begin{split}
\mathcal{M}_{\text{PC}}^{-1/2,-1/2}=&-\mathcal{M}_{\text{PC}}^{1/2,1/2},~\mathcal{M}_{\text{PV}}^{-1/2,-1/2}=\mathcal{M}_{\text{PV}}^{1/2,1/2},\\
\mathcal{M}_{\text{PV},\text{PC}}^{-1/2,1/2}=&\mathcal{M}_{\text{PV},\text{PC}}^{1/2,-1/2}=0,
\label{eq:relation2}
\end{split}
\end{equation}
for $W$-exchange diagram.

\subsection{The decay width and asymmetry parameter $\alpha$}
\label{sec2.3}

With the above preparations, we can further investigate the physical observable. The decay width can be expressed as
\begin{equation}
\Gamma[\mathcal{B}_{i}{\to}\mathcal{B}_{f}P]=8\pi^{2}\frac{kE_{f}E_{P}}{m_{i}}\frac{1}{2J_{i}+1}\sum_{s_{z}^{f},s_{z}^{i}}\Big{(}\vert\mathcal{M}_{\text{PC}}^{s_{z}^{f},s_{z}^{i}}\vert^{2}+\vert\mathcal{M}_{\text{PV}}^{s_{z}^{f},s_{z}^{i}}\vert^{2}\Big{)},
\label{eq:Br}
\end{equation}
where, $m_{i}$ is the mass of particle $\mathcal{B}_{i}$, and $E_{f}$ and $E_{P}$ are the energies of $\mathcal{B}_{f}$ and $P$, respectively. Besides, $J$ is the total spin of initial state $\mathcal{B}_{i}$. In addition, the asymmetry parameter $\alpha$ can be defined by
\begin{equation}
\alpha=\frac{2\text{Re}[(\mathcal{M}_{\text{PV}}^{-1/2,-1/2})^{*}\mathcal{M}_{\text{PC}}^{-1/2,-1/2}]}{\vert\mathcal{M}_{\text{PC}}^{-1/2,-1/2}\vert^{2}+\vert\mathcal{M}_{\text{PV}}^{-1/2,-1/2}\vert^{2}}.
\label{eq:alpha}
\end{equation}

\section{The nonrelativistic potential and baryon wave functions}
\label{sec3}

In this section, we illustrate how to obtain the wave functions of baryons. For this purpose, we employ a nonrelativistic Hamiltonian~\cite{Copley:1979wj,Pervin:2007wa,Roberts:2007ni,Yoshida:2015tia}
\begin{equation}
\mathcal{H}=\sum_{i=1,2,3}\Big{(}m_{i}+\frac{p_{i}^{2}}{2m_{i}}\Big{)}+\sum_{i<j}V_{ij},
\end{equation}
to describe the baryon system, where $m_{i}$ and $p_{i}$ are mass and momentum of $i$th quark, respectively. The nonrelativistic quark-quark interaction is given by $V_{ij} = V_{ij}^{\text{con}} + V_{ij}^{\text{short}}$ with the linear potential and the short-range potential being~\cite{Yoshida:2015tia}
\begin{equation}
\begin{split}
V_{ij}^{\text{con}}=&\frac{br_{ij}}{2}+\text{const.},\\
V_{ij}^{\text{short}}=&-\frac{2}{3}\frac{\alpha^{\text{Coul}}}{r_{ij}}
+\frac{16\pi\alpha^{ss}}{9m_{i}m_{j}}\mathbf{S}_{i}\cdot\mathbf{S}_{j}\frac{\Lambda^{2}}{4\pi r_{ij}}\text{exp}(-\Lambda r_{ij})\\
&+\frac{\alpha^{\text{so}}\big{[}1{\!}-{\!}\text{exp}(-\Lambda r_{ij})\big{]}^{2}}{3r_{ij}^{3}}
\Bigg{[}\mathbf{L}_{ij}\cdot(\mathbf{S}_{i}+\mathbf{S}_{j})\\
&\times\Bigg{(}\frac{1}{m_{i}^{2}}+\frac{1}{m_{j}^{2}}+\frac{4}{m_{i}m_{j}}\Bigg{)}+\mathbf{L}_{ij}\cdot(\mathbf{S}_{i}-\mathbf{S}_{j})\Bigg{(}\frac{1}{m_{i}^{2}}-\frac{1}{m_{j}^{2}}\Bigg{)}\Bigg{]}\\
&+\frac{2\alpha^{\text{ten}}\big{[}1{\!}-{\!}\text{exp}(-\Lambda r_{ij})\big{]}^{2}}{3m_{i}m_{j}r_{ij}^{3}}
\Bigg{[}\frac{3(\mathbf{s}_{i}\cdot\mathbf{r}_{ij})(\mathbf{s}_{j}\cdot\mathbf{r}_{ij})}{r_{ij}^{2}}-\mathbf{S}_{i}\cdot\mathbf{S}_{j}\Bigg{]},
\end{split}
\end{equation}
respectively, in which $\alpha^{\text{Coul}}=K(m_{i}+m_{j})/(m_{i}m_{j})$, and $\pmb{S}_{i}$ and $\pmb{L}_{ij}$ are spin and orbital operators, respectively. The parameters of nonrelativistic potential model are shown in Table~\ref{tab:PotentialParameters}.

\begin{table}[htbp]
\centering
\caption{The parameters of nonrelativistic potential~\cite{Yoshida:2015tia}. Besides, the masses of constituent quarks are taken as $m_{u,d}=300~\text{MeV}$, $m_{s}=510~\text{MeV}$, and $m_{c}=1750~\text{MeV}$.}
\label{tab:PotentialParameters}
\renewcommand\arraystretch{1.05}
\begin{tabular*}{80mm}{c@{\extracolsep{\fill}}ccc}
\toprule[1pt]
\toprule[0.5pt]
Parameters              &Values    &Parameters  &Values\\
\midrule[0.5pt]
$b~[\text{GeV}^{2}]$     &$0.165$   &$\alpha^{\text{ss}}$  &$1.2$\\
$\text{const.}~[\text{GeV}]$  &$-1.139$  &$\alpha^{\text{so}}=\alpha^{\text{ten}}$  &$0.077$\\
$K~[\text{MeV}]$         &$90$      &$\Lambda~[\text{fm}^{-1}]$  &$3.5$\\
\bottomrule[0.5pt]
\bottomrule[1pt]
\end{tabular*}
\end{table}

In the study of baryon spectrum, the masses and wave functions can be derived by solving Schr\"{o}dinger equation with the nonrelativisitc Hamiltonian
\begin{equation}
\mathcal{H}\vert\Psi_{J,M_{J}}\rangle=E\vert\Psi_{J,M_{J}}\rangle,
\label{eq:Schrodinger}
\end{equation}
in which the baryon wave function $\Psi_{J,M_{J}}$ is constructed as a combination of color, spin, spatial, and flavor terms:
\begin{equation}
\begin{split}
\Psi_{J,M_{J}}=&\sum_{\alpha}C^{(\alpha)}\Psi_{J,M_{J}}^{(\alpha)},\\
\Psi_{J,M_{J}}^{(\alpha)}=&\chi^{\text{color}}\Big{\{}\chi_{S,M_{S}}^{\text{spin}}\psi_{L,M_{L}}^{\text{spatial}}\Big{\}}_{J,M_{J}}\psi^{\text{flavor}}
\end{split}
\end{equation}
with $C^{(\alpha)}$ being the coefficient, and $\alpha$ representing all possible quantum numbers. The spatial wave function $\psi_{L,M_{L}}^{\text{spatial}}$ consists of both $\rho$-mode and $\lambda$-mode excitations as
\begin{equation}
\psi_{L,M_{L}}^{\text{spatial}}(\pmb{\rho},\pmb{\lambda})=\sum_{m_{\rho},m_{\lambda}}\langle{l_{\rho},m_{\rho};l_{\lambda},m_{\lambda}}\vert{L,M_{L}}\rangle\psi_{l_{\rho},m_{\rho}}(\pmb{\rho}_{c})\psi_{l_{\lambda},m_{\lambda}}(\pmb{\lambda}_{c}).
\label{eq:wavefunction2}
\end{equation}
In our calculation, the momentums of the $\rho$-mode and $\lambda$-mode are defined as
\begin{equation}
\begin{split}
\pmb{\rho}=&\frac{m_{1}\pmb{p}_{2}-m_{2}\pmb{p}_{1}}{m_{1}+m_{2}},\\
\pmb{\lambda}=&\frac{(m_{1}+m_{2})\pmb{p}_{3}-m_{3}(\pmb{p}_{1}+\pmb{p}_{2})}{m_{1}+m_{2}+m_{3}},
\end{split}
\end{equation}
respectively, which indicates that the single charmed baryon can be treated as a bound state of a light quark cluster and a charm quark, while the double charmed baryon is considered a bound state of a di-charm quark cluster and a light quark.

To expand the spatial wave functions $\psi_{l_{\rho},m_{\rho}}$ and $\psi_{l_{\lambda},m_{\lambda}}$, the infinitesimally-shift Gaussian basis~\cite{Hiyama:2003cu,Hiyama:2018ivm}
\begin{equation}
\begin{split}
\phi_{nlm}^{G}(\mathbf{r})=&\phi^{G}_{nl}(r)~Y_{lm}(\pmb{\hat{r}})\\
=&\sqrt{\frac{2^{l+2}(2\nu_{n})^{l+3/2}}{\sqrt{\pi}(2l+1)!!}}\lim_{\varepsilon\rightarrow0}
\frac{1}{(\nu_{n}\varepsilon)^l}\sum_{k=1}^{k_{\text{max}}}C_{lm,k}e^{-\nu_{n}\big{(}\mathbf{r}-\varepsilon\pmb{D}_{lm,k}\big{)}^2},
\label{eq:Gaussianbasis}
\end{split}
\end{equation}
is employed, in which the Gaussian size parameter $\nu_{n}$ in our computations follows a geometric progression sequence:~\cite{Li:2021qod,Luo:2022cun,Li:2025alu}
\begin{equation}
\nu_{n}=1/r^2_{n},~~~r_{n}=r_{min}~a^{n-1},~~~a=\Big{(}r_{max}/r_{min}\Big{)}^{\frac{1}{n_{max}-1}}
\end{equation}
with $r_{\rho_{min}}=r_{\lambda_{min}}=0.2~\text{fm}$, $r_{\rho_{max}}=r_{\lambda_{max}}=2.0~\text{fm}$, and $n_{\rho_{max}}=n_{\lambda_{max}}=6$. The Gaussian basis in momentum space can be obtained by replacing $r\to{p}$ and $\nu_{n}\to1/(4\nu_{n})$.

Finally, with the nonrelativistic potential and baryon wave function prepared, we further solve the Schr\"{o}dinger equation assisted by Gaussian expansion method~\cite{Hiyama:2003cu,Hiyama:2018ivm}:
\begin{equation}
\Big{(}T^{\alpha^{\prime},\alpha}+V^{\alpha^{\prime},\alpha}\Big{)}C^{(\alpha)}=EN^{\alpha^{\prime},\alpha}C^{(\alpha)},
\end{equation}
where the matrix elements are given by
\begin{equation}
\begin{split}
T^{\alpha^{\prime},\alpha}=&\Big{\langle}\Psi_{\mathbf{J},\mathbf{M_J}}^{(\alpha^{\prime})}\big{\vert}
\Bigg{[}\sum_{i=1,2,3}\Big{(}m_{i}+\frac{p_{i}^{2}}{2m_{i}}\Big{)}\Bigg{]}
\big{\vert}\Psi_{\mathbf{J},\mathbf{M_J}}^{(\alpha)}\Big{\rangle},\\
V^{\alpha^{\prime},\alpha}=&\Big{\langle}\Psi_{\mathbf{J},\mathbf{M_J}}^{(\alpha^{\prime})}\big{\vert}
\sum_{i<j}V_{ij}
\big{\vert}\Psi_{\mathbf{J},\mathbf{M_J}}^{(\alpha)}\Big{\rangle},\\
N^{\alpha^{\prime},\alpha}=&\Big{\langle}\Psi_{\mathbf{J},\mathbf{M_J}}^{(\alpha^{\prime})}
\big{\vert}\Psi_{\mathbf{J},\mathbf{M_J}}^{(\alpha)}\Big{\rangle}.
\label{eq:matrixelement}
\end{split}
\end{equation}
Complete numerical results for the charmed baryon mass calculations, including the corresponding coefficients of Gaussian bases, are provided in Table~\ref{tab:wavefunctions}. Additionally, we also compare our results of the masses of double charmed baryons with the experimental data and other theoretical results~\cite{Ebert:2002ig,Roberts:2007ni,Yu:2022lel,Ortiz-Pacheco:2023kjn,Shu:2024jdn}, in Table~\ref{tab:masses}.

\begin{table*}[htbp]\centering
\caption{The calculated masses and the coefficients of the Gaussian bases of spatial wave functions of charmed baryons, in which the coefficients of the Gaussian bases are presented in the third column, with the pairs $(n_{\rho},n_{\lambda})$ arranged in the sequence $\{(1,1),(1,2),\cdots,(1,n_{\lambda_{max}}),(2,1),(2,2), \cdots,(2,n_{\lambda_{max}}),$ $\cdots,(n_{\rho_{max}},1), (n_{\rho_{max}},2),\cdots,(n_{\rho_{max}},n_{\lambda_{max}})\}$.}
\label{tab:wavefunctions}
\renewcommand\arraystretch{1.05}
\begin{tabular*}{150mm}{c@{\extracolsep{\fill}}cc}
\toprule[1pt]
\toprule[0.5pt]
Charmed  &Theoretical  &\multirow{2}*{Eigenvector coefficients $C^{(\alpha)}$}\\
baryons  &values $\text{[GeV]}$  &\\
\midrule[0.5pt]
\multirow{4}*{{$\Xi_{cc}^{++}/\Xi_{cc}^{+}$}}  &\multirow{4}*{$3.688$}
&$\big{\{}-0.0010, 0.0046, -0.0089, -0.0093, -0.0033, 0.0007, -0.0140, -0.0011, 0.0103,$\\
&&$0.0269, 0.0031, -0.0004, 0.0139, -0.0509, -0.1297, -0.3786, -0.0431, 0.0042,$\\
&&$-0.0052, 0.0294, -0.0381, -0.4568, -0.1116, 0.0151, 0.0020, -0.0121, 0.0208,$\\
&&$0.0683, 0.0130, -0.0019, -0.0005, 0.0031, -0.0054, -0.0137, -0.0029, 0.0005\big{\}}$\\
\specialrule{0em}{2pt}{2pt}
\multirow{4}*{{$\Xi_{cc}^{+}(2S)$}}  &\multirow{4}*{$4.087$}
&$\big{\{}-0.0058, 0.0156, -0.0025, -0.0352, -0.0110, 0.0017, -0.0011, -0.0523, -0.0254,$\\
&&$0.1534, 0.0477, -0.0066, 0.0013, 0.0366, -0.2120, -1.4404, -0.2883, 0.0370,$\\
&&$-0.0023, -0.0271, 0.2375, 1.1653, 0.2727, -0.0334, 0.0000, 0.0169, -0.1403,$\\
&&$0.2899, 0.1584, -0.0211, -0.0005, -0.0020, 0.0288, -0.0599, -0.0281, 0.0036\big{\}}$\\
\specialrule{0em}{2pt}{2pt}
\multirow{4}*{{$\Xi_{cc}^{+}(1P,\rho)$}}  &\multirow{4}*{$3.951$}
&$\big{\{}-0.0001, 0.0017, -0.0044, -0.0077, -0.0019, 0.0003, -0.0029, -0.0033, 0.0141,$\\
&&$0.0306, 0.0050, -0.0008, 0.0047, -0.0151, -0.0803, -0.2901, -0.0467, 0.0056,$\\
&&$-0.0009, 0.0061, 0.0042, -0.5953, -0.1816, 0.0240, 0.0010, -0.0059, 0.0136,$\\
&&$0.0749, 0.0120, -0.0016, -0.0003, 0.0017, -0.0039, -0.0159, -0.0026, 0.0004\big{\}}$\\
\specialrule{0em}{2pt}{2pt}
\multirow{4}*{{$\Xi_{cc}^{+}(1P,\lambda)$}}  &\multirow{4}*{$4.145$}
&$\big{\{}0.0007, -0.0013, -0.0004, -0.0107, -0.0038, 0.0005, -0.0024, 0.0052, 0.0015,$\\
&&$0.0211, 0.0050, -0.0006, 0.0022, -0.0178, -0.0116, -0.2964, -0.1083, 0.0145,$\\
&&$0.0027, -0.0061, 0.0097, -0.5237, -0.2485, 0.0342, -0.0006, -0.0000, 0.0069,$\\
&&$0.0592, 0.0211, -0.0030, 0.0002, -0.0001, -0.0016, -0.0120, -0.0046, 0.0007\big{\}}$\\
\specialrule{0em}{2pt}{2pt}
\multirow{4}*{{$\Xi_{cc}^{+}(2P,\rho)$}}  &\multirow{4}*{$4.303$}
&$\big{\{}0.0013, -0.0044, -0.0017, 0.0327, 0.0100,-0.0013, 0.0007, 0.0142, 0.0156,$\\
&&$-0.1531, -0.0479, 0.0063, -0.0027, 0.0022, 0.0767, 1.166, 0.2915, -0.0376,$\\
&&$0.0009, 0.0042, -0.0630, -0.7838, -0.2406, 0.0287, 0.0031, -0.0200, 0.1008,$\\
&&$-0.3323, -0.2109, 0.0281, -0.0004, 0.0034, -0.0206, 0.0669, 0.0376, -0.0050\big{\}}$\\
\specialrule{0em}{2pt}{2pt}
\multirow{4}*{{$\Xi_{cc}^{+}(2P,\lambda)$}}  &\multirow{4}*{$4.496$}
&$\big{\{}-0.0001, -0.0017, 0.0001, 0.0396, 0.0189, -0.0026, 0.0016, 0.0049,$\\
&&$0.0021, -0.1697, -0.0835, 0.0115, -0.0062, 0.0199, 0.0095, 1.3023, 0.5682, -0.0779,$\\
&&$0.0041, -0.0110, -0.0560, -0.7941, -0.3187, 0.0422, 0.0021, -0.0131, 0.0916,$\\
&&$-0.4780, -0.3387, 0.0449, -0.0001, 0.0012, -0.0151, 0.0862, 0.0579, -0.0077\big{\}}$\\
\specialrule{0em}{2pt}{2pt}
\multirow{4}*{{$\Xi_{c}^{\bar{3},+}$}}  &\multirow{4}*{$2.500$}
&$\big{\{}-0.0005, -0.0006, -0.0104, -0.0137, 0.0009, -0.0002, -0.0059, 0.0089, -0.0168,$\\
&&$-0.0243, 0.0028, -0.0007, 0.0024, -0.0353, -0.0271, -0.0476, 0.0035, -0.0008,$\\
&&$-0.0030, 0.0266, -0.3020, -0.3543, 0.0370, -0.0083, -0.0042, 0.0200, -0.0665,$\\
&&$-0.3414, 0.0273, -0.0064, 0.0008, -0.0036, 0.0100, 0.0351, -0.0046, 0.0011\big{\}}$\\
\specialrule{0em}{2pt}{2pt}
\multirow{4}*{{$\Xi_{c}^{6,+}$}}  &\multirow{4}*{$2.603$}
&$\big{\{}0.0006, -0.0015, 0.0021, 0.0038, -0.0009, 0.0002, -0.0094, 0.0181, -0.0181,$\\
&&$-0.0128, 0.0017, -0.0004, 0.0043, -0.0575, 0.0569, 0.0325, -0.0037, 0.0008,$\\
&&$-0.0003, 0.0193, -0.3614, -0.2730, 0.0295, -0.0064, -0.0072, 0.0347, -0.1110,$\\
&&$-0.4803, 0.0435, -0.0101, 0.0014, -0.0063, 0.0171, 0.0439, -0.0072, 0.0017\big{\}}$\\
\toprule[0.5pt]
\toprule[1pt]
\end{tabular*}
\end{table*}

\begin{table*}[htbp]\centering
\caption{The comparison of theoretical and experimental values of masses of double charmed baryons, in which the values are presented in the units of MeV.}
\label{tab:masses}
\renewcommand\arraystretch{1.05}
\begin{tabular*}{120mm}{c@{\extracolsep{\fill}}ccccccc}
\toprule[1pt]
\toprule[0.5pt]
States                 &This work &Ref.~\cite{Ebert:2002ig} &Ref.~\cite{Roberts:2007ni} &Ref.~\cite{Yu:2022lel} &Ref.~\cite{Ortiz-Pacheco:2023kjn} &Ref.~\cite{Shu:2024jdn} &Expt.~\cite{ParticleDataGroup:2024cfk}\\
\midrule[0.5pt]
$\Xi_{cc}(1S)$         &$3688$   &$3620$                  &$3676$                    &$3640$                &$3619$                           &$3621$       &$3621.6\pm0.4$\\
$\Xi_{cc}(2S)$         &$4087$   &$3910$                  &$4029$                    &$4069$                &                                 &             &\\
$\Xi_{cc}(1P,\rho)$    &$3951$   &                        &$3910$                    &$3932$                &$3823$                           &$3883$       &\\
$\Xi_{cc}(1P,\lambda)$ &$4145$   &                        &$4074$                    &                      &$4048$                           &$4071$       &\\
\bottomrule[0.5pt]
\bottomrule[1pt]
\end{tabular*}
\end{table*}

\section{Numerical results}
\label{sec4}

With the above preparations, we can subsequently compute the decay amplitudes numerically. In our calculation, the CKM matrix elements are adopted as the Wolfenstein parameterization as
\begin{equation*}
V_{ud}=1-\lambda^{2}/2,~~V_{cs}=1-\lambda^{2}/2
\end{equation*}
with $\lambda=0.22501$~\cite{ParticleDataGroup:2024cfk}. The life time of double charmed baryon $\Xi_{cc}^{++}$ is chosen as~\cite{ParticleDataGroup:2024cfk}
\begin{equation*}
\tau_{\Xi_{cc}^{++}}=(2.56\pm0.27)~\times10^{-13}s.
\end{equation*}
Besides, the masses of ground states, $\Xi_{cc}^{++,+}$ and $\Xi_{c}^{(\prime)+}$, are chosen from Particle Data Group~\cite{ParticleDataGroup:2024cfk}, while the ones of excited double charmed baryons are chosen from Table~\ref{tab:wavefunctions}.

At the first, we calculate the decay amplitudes $\mathcal{M}_{\text{PC},\text{PV}}^{-1/2,-1/2}$ of $T$, $C^{\prime}$ and $W$-exchange diagrams, using Eqs.~\eqref{eq:amplitudes1} and \eqref{eq:amplitudes2}. Our numerical results are displayed in Table~\ref{tab:amplitudes}, in which the differences of amplitudes between $\Xi_{cc}^{++}\to\Xi_{c}^{+}\pi^{+}$ and $\Xi_{cc}^{++}\to\Xi_{c}^{\prime+}\pi^{+}$ processes originate from the different momentum $\pmb{k}$. {Obviously, the contribution from $\Xi_{cc}(2S)$ pole is significantly smaller than that from the ground state $\Xi_{cc}(1S)$ pole. Besides, the contributions from 2P states are also significantly smaller than that from 1P states. This guarantees that the contributions from highly excited states can be ignored safely.}

In addition, the amplitudes $\mathcal{M}_{\text{PC},\text{PV}}^{1/2,1/2}$ can be obtained with the relations in Eqs.~\eqref{eq:relation1} and \eqref{eq:relation2}. Finally, the total amplitudes $\mathcal{M}_{\text{PC}}^{-1/2,-1/2}$ and $\mathcal{M}_{\text{PV}}^{-1/2,-1/2}$ can be obtained through summation over all contributing diagrams:
\begin{equation}
\begin{split}
\mathcal{M}_{\text{PC}}^{-1/2,-1/2}=&\mathcal{M}_{\text{PC},T}^{-1/2,-1/2}+\mathcal{M}_{\text{PC},C^{\prime}}^{-1/2,-1/2}\\
&+\mathcal{M}_{\text{PC},\Xi_{cc}^{+}(1S)-pole}^{-1/2,-1/2}+\mathcal{M}_{\text{PC},\Xi_{cc}^{+}(2S)-pole}^{-1/2,-1/2},\\
\mathcal{M}_{\text{PV}}^{-1/2,-1/2}=&\mathcal{M}_{\text{PV},T}^{-1/2,-1/2}+\mathcal{M}_{\text{PV},C^{\prime}}^{-1/2,-1/2}\\
&+\mathcal{M}_{\text{PV},\Xi_{cc}^{+}(1P,\rho)-pole}^{-1/2,-1/2}+\mathcal{M}_{\text{PV},\Xi_{cc}^{+}(1P,\lambda)-pole}^{-1/2,-1/2},
\end{split}
\end{equation}
and the total amplitudes $\mathcal{M}_{\text{PC}}^{1/2,1/2}$ and $\mathcal{M}_{\text{PV}}^{1/2,1/2}$ can be obtained similarly.

\begin{table*}[htbp]\centering
\caption{The amplitudes $\mathcal{M}_{\text{PC},\text{PV}}^{-1/2,-1/2}$ (in the units of $10^{-4}G_{F}~\text{GeV}^{3/2}$) of $\Xi_{cc}^{++}\to\Xi_{c}^{+}\pi^{+}$ and $\Xi_{cc}^{++}\to\Xi_{c}^{\prime+}\pi^{+}$ processes.}
\label{tab:amplitudes}
\renewcommand\arraystretch{1.05}
\begin{tabular*}{150mm}{c@{\extracolsep{\fill}}ccccc}
\toprule[1pt]
\toprule[0.5pt]
\specialrule{0em}{1pt}{1pt}
& &\multicolumn{2}{c}{{\normalsize $\Xi_{cc}^{++}\to\Xi_{c}^{+}\pi^{+}$ process}} &\multicolumn{2}{c}{{\normalsize $\Xi_{cc}^{++}\to\Xi_{c}^{\prime+}\pi^{+}$ process}}\\
\specialrule{0em}{1pt}{1pt}
\midrule[0.5pt]
\specialrule{0em}{0.5pt}{0.5pt}
& &$\mathcal{M}[\Xi_{cc}^{++}\to\Xi_{c}^{\bar{3},+}\pi^{+}]$  &$\mathcal{M}[\Xi_{cc}^{++}\to\Xi_{c}^{6,+}\pi^{+}]$  &$\mathcal{M}[\Xi_{cc}^{++}\to\Xi_{c}^{\bar{3},+}\pi^{+}]$  &$\mathcal{M}[\Xi_{cc}^{++}\to\Xi_{c}^{6,+}\pi^{+}]$\\
\specialrule{0em}{0.5pt}{0.5pt}
\midrule[0.5pt]
\multirow{1}*{\shortstack{$\mathcal{M}_{\text{PC}}^{-1/2,-1/2}$}}
&$T-diagram$  &$-9.51$  &$-26.27$  &$-9.79$  &$-27.12$\\
&$C^{\prime}-diagram$  &$0$  &$-2.58$  &$0$  &$-2.71$\\
&$\Xi_{cc}^{+}(1S)-pole$  &$-31.20$  &$0$  &$-39.55$  &$0$\\
&$\Xi_{cc}^{+}(2S)-pole$  &$-0.85$  &$0$  &$-0.87$  &$0$\\
&$\text{total}$  &$-41.56$  &$-28.85$  &$-50.21$  &$-29.83$\\
\specialrule{0em}{2pt}{2pt}
\multirow{1}*{\shortstack{$\mathcal{M}_{\text{PV}}^{-1/2,-1/2}$}}
&$T-diagram$  &$14.36$  &$8.01$  &$16.06$  &$8.99$\\
&$C^{\prime}-diagram$  &$6.95$  &$0$  &$7.94$  &$0$\\
&$\Xi_{cc}^{+}(1P,\rho)-pole$  &$0$  &$0$  &$0$  &$0$\\
&$\Xi_{cc}^{+}(1P,\lambda)-pole$  &$-2.17$  &$0$  &$-2.58$  &$0$\\
&${\Xi_{cc}^{+}(2P,\rho)-pole}$  &$0$  &$0$  &$0$  &$0$\\
&${\Xi_{cc}^{+}(2P,\lambda)-pole}$  &$-0.08$  &$0$  &$-0.08$  &$0$\\
&$\text{total}$  &$19.06$  &$8.01$  &$21.34$  &$8.99$\\
\bottomrule[0.5pt]
\bottomrule[1pt]
\end{tabular*}
\end{table*}

\begin{figure*}[htbp]\centering
  \begin{tabular}{cc}
  \includegraphics[width=68mm]{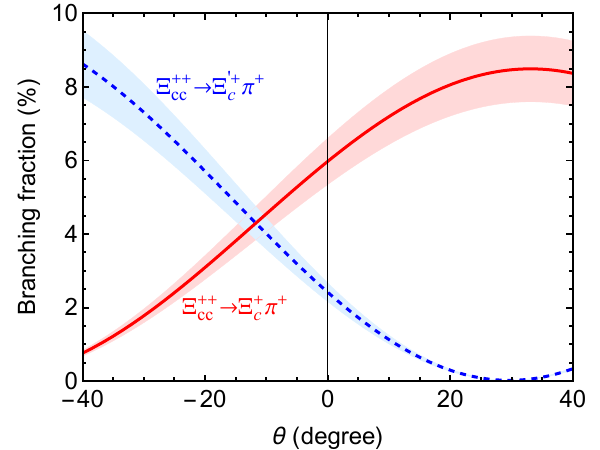}
  \includegraphics[width=68mm]{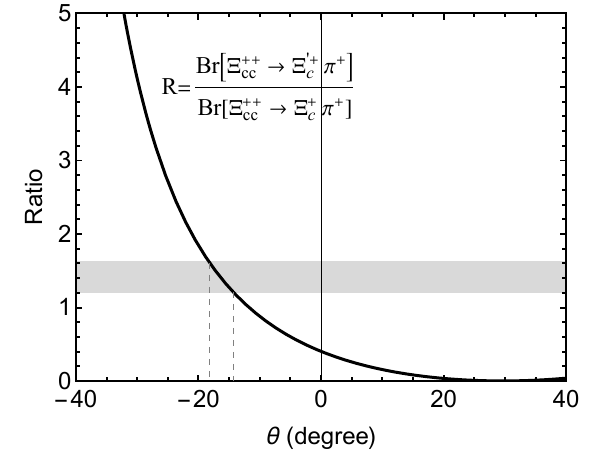}
  \end{tabular}
  \caption{The $\theta$ dependence of branching fractions (left panel) of $\Xi_{cc}^{++}\to\Xi_{c}^{(\prime)+}\pi^{+}$ processes and their ratio (right panel).}
\label{fig:Brratio}
\end{figure*}

With the obtained decay amplitudes, we further investigate the branching fractions of $\Xi_{cc}^{++}\to\Xi_{c}^{(\prime)+}\pi^{+}$ processes by using Eq.~\eqref{eq:Br}. In the left panel of Fig.~\ref{fig:Brratio}, we present the dependence of branching fractions on the mixing angle $\theta$. It is clear that the predicted branching fractions are at the order of magnitude of $10^{-2}$, and show highly sensitive to the mixing angle. In order to restrict the mixing angle, we utilize the ratio of branching fractions $R=1.41\pm0.20$~\cite{LHCb:2022rpd}, which was measured by the LHCb Collaboration. The $\theta$ dependence of this ratio is shown in the right panel of Fig.~\ref{fig:Brratio}, where the black curve is our result and the light gray region corresponds to experimental value. Our result shows this ratio is highly sensitive to the mixing angle $\theta$. Obviously, the experimental value can be well reproduced for the mixing angle range $\theta\in(-18.2^{\circ},-14.3^{\circ})$, and the center value can be obtained with $\theta=-16.4^{\circ}$. Within the established range, the absolute branching fractions of $\Xi_{cc}^{++}\to\Xi_{c}^{(\prime)+}\pi^{+}$ processes are determined as
\begin{equation}
\begin{split}
\mathcal{B}[\Xi_{cc}^{++}\to\Xi_{c}^{+}\pi^{+}]=&(3.0\sim4.3)\%,\\
\mathcal{B}[\Xi_{cc}^{++}\to\Xi_{c}^{\prime+}\pi^{+}]=&(4.2\sim6.0)\%,
\end{split}
\end{equation}
which is significantly larger than the estimation
\begin{equation}
\mathcal{B}[\Xi_{cc}^{++}\to\Xi_{c}^{+}\pi^{+}]_{\text{expt}}\approx(1.33\pm0.74)\%,
\end{equation}
by Refs.~\cite{Liu:2022igi,Zeng:2022egh}.

{The nonleptonic decay $\Xi_{cc}^{++}\to\Xi_{c}^{+}\pi^{+}$ was first observed by LHCb Collaboration in 2018 with the integrated luminosity of $1.7~\text{fb}^{-1}$~\cite{LHCb:2018pcs}. Subsequently, in 2022, the decay mode $\Xi_{cc}^{++}\to\Xi_{c}^{\prime+}\pi^{+}$ was also observed by LHCb with the integrated luminosity of $5.4~\text{fb}^{-1}$~\cite{LHCb:2022rpd}. According to Ref.~\cite{LHCb:2022rpd}, the yields of decay modes $\Xi_{cc}^{++}\to\Xi_{c}^{(\prime)+}\pi^{+}$ were only several hundred, which indicates that the measurement of absolute branching fraction remains challenging. However, with the accumulation of experimental data in LHCb experiment, the measurement is becoming possible.}

Compared to nonleptonic processes, semileptonic decays of $\Xi_{c}^{(\prime)}$ provides a more clearer platform to determine the $\Xi_{c}-\Xi_{c}^{\prime}$ mixing effect. Recently, the analysis of $\Xi_{c}^{0}\to\Xi^{-}e^{+}\nu_{e}$ based on QCD sum rule suggested a small mixing angle as $\theta=(1.2-2.8)^{\circ}$~\cite{Sun:2023noo}. A small mixing angle as $\theta=(1.2\pm0.1)^{\circ}$ is also obtained in the analysis of $\Xi_{c}$ mass spectroscopy by lattice QCD~\cite{Liu:2023feb,Liu:2023pwr}. In this work, when taking $\theta=0^{\circ}$, the branching fractions of $\Xi_{cc}^{++}\to\Xi_{c}^{(\prime)+}\pi^{+}$ are predicted as
\begin{equation*}
\begin{split}
\mathcal{B}[\Xi_{cc}^{++}\to\Xi_{c}^{+}\pi^{+}]=&(6.00\pm0.60)\%,\\
\mathcal{B}[\Xi_{cc}^{++}\to\Xi_{c}^{\prime+}\pi^{+}]=&(2.42\pm0.25)\%.
\end{split}
\end{equation*}
The precise measurements of absolute branching fractions by the ongoing LHCb and Belle II experiments will be helpful for further confirming the mixing angle.

\begin{figure}[htbp]\centering
  \includegraphics[width=68mm]{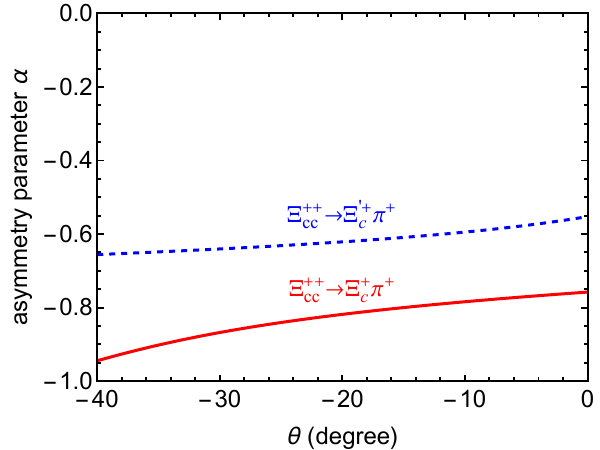}\\
  \caption{The $\theta$ dependence of asymmetry parameters $\alpha$ of $\Xi_{cc}^{++}\to\Xi_{c}^{(\prime)+}\pi^{+}$ processes.}
  \label{fig:alpha}
\end{figure}

Additionally, we also investigate the dependence of asymmetry parameter $\alpha$ on the mixing angle $\theta$, by using Eq.~\eqref{eq:alpha}. In Fig.~\ref{fig:alpha}, we show the $\theta$ dependence of the $\alpha$. In the region of mixing angle $\theta\in(-18.2^{\circ},-14.3^{\circ})$, the asymmetry parameters are predicted as
\begin{equation*}
\begin{split}
\alpha[\Xi_{cc}^{++}\to\Xi_{c}^{+}\pi^{+}]=&(-0.80\sim-0.81),\\
\alpha[\Xi_{cc}^{++}\to\Xi_{c}^{\prime+}\pi^{+}]=&(-0.61\sim-0.62),
\end{split}
\end{equation*}
and it shows that the asymmetry parameters are less sensitive to the mixing angle. {In experiment, the decay asymmetry parameter $\alpha$ can be extracted by fitting the angular distribution of corresponding decay mode. At this moment, due to the small sample size in LHCb experiment, the determination of $\alpha$ is impossible. We are looking for more experimental data, and expect the asymmetry parameter can be measured.} Future measurements of these observables will contribute to validate our predictions.

As is well known, the decay amplitude of two-body nonleptonic process $\mathcal{B}_{i}\to\mathcal{B}_{f}+P$ can be parametrized as
\begin{equation}
\mathcal{M}[\mathcal{B}_{i}\to\mathcal{B}_{f}+P]=i\bar{u}_{f}(A-B\gamma_{5})u_{i},
\end{equation}
where, $A$ and $B$ are $\text{PV}$ and $\text{PC}$ amplitudes, respectively. They can convert to $\mathcal{M}_{\text{PC,PV}}^{-1/2,-1/2}$ via the relations as
\begin{equation}
\begin{split}
\mathcal{M}_{\text{PV}}^{-1/2,-1/2}=&\frac{i}{4\pi}\sqrt{\frac{m_{f}}{{\pi}E_{f}E_{P}}}\sqrt{\frac{E_{f}+m_{f}}{2m_{f}}}\chi_{f}^{\dagger}\chi_{i}A,\\
\mathcal{M}_{\text{PC}}^{-1/2,-1/2}=&\frac{i}{4\pi}\sqrt{\frac{m_{f}}{{\pi}E_{f}E_{P}}}\sqrt{\frac{E_{f}+m_{f}}{2m_{f}}}\chi_{f}^{\dagger}\frac{\pmb{\sigma}\cdot\pmb{P}_{f}}{E_{f}+m_{f}}\chi_{i}B,
\end{split}
\end{equation}
where $\chi_{f,i}$ are the spin wave functions of charmed baryons. In Table~\ref{tab:CompareAmplitudes}, we compare the factorizable amplitudes $A^{\text{fac}}$ and $B^{\text{fac}}$, and nonfactorizable amplitudes $A^{\text{nf}}$ and $B^{\text{nf}}$, obtained by different theoretical works~\cite{Gutsche:2018msz,Cheng:2020wmk,Liu:2022igi,Zeng:2022egh,Shi:2022kfa}. The nonfactorizable amplitude $B^{\text{nf}}$ of $\Xi_{cc}^{++}\to\Xi_{c}^{\prime+}\pi^{+}$ decay is greatly enhanced due to the mixing effect. Comprehensive experimental measurements of concerned physical observables will be helpful for validating different theoretical models.

\begin{table*}[htbp]\centering
\caption{The comparison of factorizable and nonfactorizable amplitudes, as well as the branching fractions (only the center value) and asymmetry parameters $\alpha$, of $\Xi_{cc}^{++}\to\Xi_{c}^{+}\pi^{+}$ (up panel) and $\Xi_{cc}^{++}\to\Xi_{c}^{\prime+}\pi^{+}$ processes (bottom panel) by various theoretical works. Here, the amplitudes are given in the units of $10^{-2}G_{F}\text{GeV}^{-2}$.}
\label{tab:CompareAmplitudes}
\renewcommand\arraystretch{1.05}
\begin{threeparttable}
\begin{tabular*}{150mm}{l@{\extracolsep{\fill}}cccccc}
\toprule[1pt]
\toprule[0.5pt]
\specialrule{0em}{0.5pt}{0.5pt}
$\Xi_{cc}^{++}\to\Xi_{c}^{+}\pi^{+}$ process   &$A^{\text{fac}}$   &$A^{\text{nf}}$   &$B^{\text{fac}}$   &$B^{\text{nf}}$  &$\mathcal{B}$  &$\alpha$\\
\specialrule{0em}{0.5pt}{0.5pt}
\midrule[0.5pt]
\multirow{1}*{\text{This work}}  &  &  &  &  &  &\\
\quad $\theta=0$  &$3.21$  &$1.05$  &$-11.30$  &$-38.09$  &$6.00\%$  &$-0.76$\\
\quad $\theta=-16.4^{\circ}$  &$2.57$    &$1.01$    &$-2.03$    &$-35.65$  &$3.60\%$  &$-0.80$\\
CCQM~\cite{Gutsche:2018msz}   &$-7.72$    &$10.92$    &$12.33$    &$-17.61$  &$0.70\%$  &\\
pole model~\cite{Cheng:2020wmk}   &$7.40$    &$-10.79$    &$-15.06$    &$18.91$  &$0.69\%$  &$-0.41$\\
\multirow{1}*{pole model~\cite{Liu:2022igi}$~^{[a]}$} &  &  &  &  &  &\\
\quad SB($\theta=-24.7^{\circ}$)  &$4.83$  &$-9.99$  &$5.16$  &$13.6$  &$2.24\%$  &$-0.93$\\
\quad HB($\theta=24.7^{\circ}$)  &$7.08$  &$-20.3$  &$-22.1$  &$33.0$  &$10.3\%$  &$-0.30$\\
pole model~\cite{Zeng:2022egh}$~^{[b]}$   &$5.5$    &$-9.6$    &$-9.6$    &$16.8$  &$1.20\%$  &$-0.78$\\
LCSR~\cite{Shi:2022kfa}   &    &$-16.67\pm1.41$    &    &$20.47\pm3.89$  &  &\\
\midrule[0.5pt]
\specialrule{0em}{0.5pt}{0.5pt}
$\Xi_{cc}^{++}\to\Xi_{c}^{\prime+}\pi^{+}$ process   &$A^{\text{fac}}$   &$A^{\text{nf}}$   &$B^{\text{fac}}$   &$B^{\text{nf}}$  &$\text{Br}$  &$\alpha$\\
\specialrule{0em}{0.5pt}{0.5pt}
\midrule[0.5pt]
\multirow{1}*{\text{This work}}  &  &  &  &  &  &\\
\quad $\theta=0$  &$1.92$  &$0$  &$-34.75$  &$-3.57$  &$2.42\%$  &$-0.55$\\
\quad $\theta=-16.4^{\circ}$  &$2.81$    &$0.32$    &$-36.88$    &$-18.05$  &$5.10\%$  &$-0.61$\\
CCQM~\cite{Gutsche:2018msz}   &$-4.12$    &$-0.11$    &$35.75$    &$1.30$  &$3.19\%$  &\\
pole model~\cite{Cheng:2020wmk}   &$4.49$    &$-0.04$    &$-48.50$    &$0.06$  &$4.65\%$  &$-0.84$\\
\multirow{1}*{pole model~\cite{Liu:2022igi}$~^{[a]}$} &  &  &  &  &  &\\
\quad SB($\theta=-24.7^{\circ}$)  &$7.38$  &$-4.82$  &$-51.0$  &$7.26$  &$3.25\%$  &$-0.63$\\
\quad HB($\theta=24.7^{\circ}$)  &$0.61$  &$9.65$  &$-28.1$  &$-17.4$  &$8.91\%$  &$-0.96$\\
pole model~\cite{Zeng:2022egh}$~^{[b]}$   &$2.8$    &$0$    &$-27.6$    &$0$  &$2.16\%$  &$-0.89$\\
LCSR~\cite{Shi:2022kfa}   &    &$-0.83\pm0.28$    &    &$8.86\pm2.16$  &  &\\
\bottomrule[0.5pt]
\bottomrule[1pt]
\end{tabular*}
\begin{tablenotes}
\footnotesize
\item[a] The results denoted by SB and HB are obtained without and with removing the center-of-mass motion. The numerical results of amplitudes are quoted from Ref.~\cite{Zeng:2022egh}.
\item[b] The results are obtained with $\alpha_{\rho1}=0.53~\text{GeV}$ and $\alpha_{\rho2}=0.17~\text{GeV}$.
\end{tablenotes}
\end{threeparttable}
\end{table*}

\section{Summary}
\label{sec5}

Both the nonperturbative nature of QCD and the complexity of weak decay mechanisms make the study of nonleptonic decays of charmed baryons remain challenging. In this work, we investigate the double charmed baryon two-body nonleptonic decays $\Xi_{cc}^{++}\to\Xi_{c}^{(\prime)+}\pi^{+}$ within the framework of NRQM.

To reproduce the LHCb Collaboration's reported ratio $R=\mathcal{B}[\Xi_{cc}^{++}\to\Xi_{c}^{\prime+}\pi^{+}]/\mathcal{B}[\Xi_{cc}^{++}\to\Xi_{c}^{+}\pi^{+}]=1.41\pm0.17\pm0.10$, both $\Xi_{c}-\Xi_{c}^{\prime}$ mixing effects and nonfactorizable amplitudes are taken into account. Moreover, in numerical calculations of decay amplitudes, the exact spatial wave functions obtained by solving the Schr\"{o}dinger equation with a nonrelativistic potential assisted by the Gaussian expansion method are adopted to reduce uncertainties from baryon wave function choices, instead of relying on oversimplified Gaussian-type wave functions. With the support from charmed baryon spectroscopy, our strategy reduces reliance on arbitrary baryon wave functions in calculating decay amplitudes, thereby lowering the model-dependent uncertainties.

With the obtained decay amplitudes, we further investigate the branching fractions. Our results show that, within the mixing angle $\theta\in(-18.2^{\circ},-14.3^{\circ})$, the ratio $R$ can be well reproduced. Next, the absolute branching fractions can be determined as $\mathcal{B}[\Xi_{cc}^{++}\to\Xi_{c}^{+}\pi^{+}]=(3.2\sim4.3)\%$ and $\mathcal{B}[\Xi_{cc}^{++}\to\Xi_{c}^{\prime+}\pi^{+}]=(4.2\sim6.0)\%$, respectively. Moreover, we predict the asymmetry parameters as $\alpha[\Xi_{cc}^{++}\to\Xi_{c}^{+}\pi^{+}]=(-0.80\sim-0.81)$ and $\alpha[\Xi_{cc}^{++}\to\Xi_{c}^{\prime+}\pi^{+}]=(-0.61\sim-0.62)$, which shows less sensitive to the mixing angle, and the measurements at ongoing LHCb and Belle II experiments will contribute to validate our predictions.

\section*{ACKNOWLEDGMENTS}

Y.-S. Li is supported by the National Nature Science Foundation of China under Grant No. 12447155, and by the Postdoctoral Fellowship Program of CPSF under Grant No. GZC20240056.

\end{document}